\RequirePackage{lineno}
\documentclass[aps,prl,superscriptaddress,altaffilletter,floatfix,preprintnumbers,amsmath,
amssymb,amsbsy,nofootinbib,nobibnotes,twocolumn,showpacs,tighten]{revtex4-1}

\usepackage{graphicx,color,units,bm,subfigure}
\newcommand{\svnid}[1]{ } 

\begin{document}

\title{Digging deeper: Observing primordial gravitational waves below  the binary black hole produced stochastic background}
\date{\today}
\author{ T.~Regimbau}
\email{regimbau@oca.eu}
\affiliation {Artemis, Universit\'e C\^ote d'Azur, CNRS, Observatoire C\^ote d'Azur, CS 34229, Nice cedex 4, France}
\author{M.~Evans}
\affiliation {LIGO, Massachusetts Institute of Technology, Cambridge, MA 02139, USA }
\author{N.~Christensen}
\email{nelson.christensen@oca.eu}
\affiliation {Artemis, Universit\'e C\^ote d'Azur, CNRS, Observatoire C\^ote d'Azur, CS 34229, Nice cedex 4, France}
\affiliation {Physics and Astronomy, Carleton College, Northfield MN 55057, USA}
\author{E.~Katsavounidis}
\affiliation {LIGO, Massachusetts Institute of Technology, Cambridge, MA 02139, USA }
\author{B.~Sathyaprakash}
\email{bss25@psu.edu}
\affiliation {Department of Physics, The Pennsylvania State University, University Park, PA 16802, USA}
\affiliation {School of Physics and Astronomy, Cardiff University, Cardiff, CF24 3AA, UK}
\author{S.~Vitale}
\affiliation {LIGO, Massachusetts Institute of Technology, Cambridge, MA 02139, USA }

\begin{abstract}
The merger rate of black hole binaries inferred from the detections in the first Advanced LIGO science run,
implies that a stochastic background produced by a cosmological population of
mergers will likely mask the primordial gravitational-wave background. Here we
demonstrate that the next generation of ground-based detectors, such as the
Einstein Telescope and Cosmic Explorer, will be able to observe binary black
hole mergers throughout the universe with sufficient efficiency that the
confusion background can potentially be subtracted to observe the primordial background at 
the level of $\Omega_{\mathrm{GW}} \simeq 10^{-13}$ after five years of observation.
\end{abstract}


\maketitle

\noindent
{\em Introduction} --- 
According to various cosmological scenarios, we are bathed in a stochastic primordial
gravitational-wave background (PGWB) produced in the very early stages of the
Universe.  Proposed theoretical models include the amplification of vacuum
fluctuations during inflation\cite{1975JETP...40..409G, 1993PhRvD..48.3513G,
1979ZhPmR..30..719S}, pre-Big Bang models \cite{1993APh.....1..317G,
1997PhRvD..55.3330B, 2010PhRvD..82h3518D}, cosmic (super)strings
\cite{2005PhRvD..71f3510D, 2007PhRvL..98k1101S, 2010PhRvD..81j4028O,
2012PhRvD..85f6001R} or phase transitions \cite{2008PhRvD..77l4015C,
2009PhRvD..79h3519C, 2009JCAP...12..024C}. The detection of a primordial
background would have a profound impact on our understanding of the evolution of
the Universe, as it represents a unique window on its first instants, up to the
limits of the Planck era, and on the physical laws that apply at the highest
energy scales. 

In addition to the PGWB, an astrophysical background is expected to
result from the superposition of a large number of unresolved sources since the
beginning of stellar activity (see \cite{2011RAA....11..369R}, for a review of
different sources that could produce an astrophysical background).   
The astrophysical background potentially contains a
wealth of information about the history and evolution of a population of 
point sources, but it is a {\it confusion noise background} that is 
detrimental to the observation of the PGWB. In this Letter we show that at the 
sensitivity levels envisaged for third generation detectors such as the Einstein 
Telescope (ET) \cite{cqg.27.194002.10} and Cosmic Explorer (CE) \cite{2016arXiv160708697A}, it will be possible to detect most of the sources, giving hope that the confusion background can be subtracted from the data, enabling the study of a PGWB.
This problem is similar to the one investigated by \cite{2006PhRvD..73d2001C,2008PhRvD..77l3010H} in the context of the Big Bang Observer.



On September $14^{th},$ 2015, Advanced LIGO~\cite{2016PhRvL.116m1103A,cqg.27.084006.10,
cqg.32.074001.15} directly detected gravitational-waves (GWs) from the collision
of two stellar-mass black holes at a redshift of $z \sim 0.1$
(GW150914) \cite{gw150914,gw150914PE}. The inferred component masses of $m_1=
36$\, M$_\odot$ and $m_2=29$\, M$_\odot$ are larger than those of candidate
black holes in X-ray binaries inferred from reliable dynamical measurements 
\cite{gw150914astro}. This first detection suggests the existence of a population of 
black holes with relatively large masses, that might have formed in low-metallicity 
stellar environments \cite{gw150914astro}, either through the evolution of an isolated
massive binary in a galaxy \cite{2010ApJ...715L.138B} or through mass segregation and dynamical
interactions in a dense globular system \cite{2015PhRvL.115e1101R}. 

LIGO discoveries during the first observing run included a high-confidence
($>5\sigma$) detection of a second merger event GW151226 and a marginal event of
lower significance $(<2\sigma)$ LVT151012, both believed to be binary black hole
(BBH) mergers. GW151226 resulted from the merger of black holes of mass
$m_1=14.2$\, M$_\odot$ and $m_2=7.5$\, M$_\odot$ \cite{2016arXiv160604856T}, and
LVT151012 is believed to have resulted from the merger of black holes of mass
$m_1=23$\,M$_\odot$ and $m_2=13$\, M$_\odot$ \cite{2016arXiv160604856T}. These observations indicate 
that many more detections will occur in the future and have provided the
tightest constraints on the rate of such events \cite{gw150914rate}.  

Besides the loudest and closest events that can be detected individually by the Advanced LIGO - Advanced Virgo network, the population of undetected sources at larger redshift is expected to create a significant astrophysical background \cite{gw150914sgwb}. The background from the
population of binary neutron stars (BNSs) and BBHs has been investigated by many authors in the past
(see~\cite{2011RAA....11..369R, 2011ApJ...739...86Z, 2011PhRvD..84h4004R,
2011PhRvD..84l4037M, 2012PhRvD..85j4024W, 2013MNRAS.431..882Z,
2015A&A...574A..58K}, for the most recent papers), who suggested that Advanced
LIGO and Advanced Virgo had a realistic chance of detecting this background
after a few years of operation with the standard cross-correlation method, even if this background is nor continuous (no overlap of the sources) or Gaussian \cite{2015PhRvD..92f3002M}.

In Ref.\,\cite{gw150914sgwb} the LIGO and Virgo collaborations calculated the
contribution to the stochastic background from BBHs with the same masses as
GW150914. Taking into account the statistical uncertainty in the rate, they
found that the stochastic signal could be detected, in the most optimistic case, even
before the design sensitivity of the instruments is reached, but more likely
after a few years of their operation at design sensitivity. It was also shown
that lower mass systems that are too faint to be detected individually could add
a significant contribution to the background. Following this first paper, other
authors have investigated the implication of GW150914 for the confusion
background, including models of metallicity evolution with redshift and mass
distributions \cite{2016arXiv160404288D, 2016arXiv160502146N}, and arrived at
the same conclusion: the background from BBHs is likely to be higher than
previously expected and may dominate over the primordial background.

In this paper, we use Monte Carlo simulations to calculate the confusion
background from BBHs observed by networks of ground-based detectors. We study
the potential reduction in the level of this background as more BBH signals are detected, 
and can be subtracted from the data, because of the improved sensitivity of ET \cite{cqg.27.194002.10} and 
CE \cite{2016arXiv160708697A} compared to advanced detectors. We show that 
the confusion background of astrophysically produced GWs can be significantly 
reduced, paving the way to observe the primordial background. 
We do not investigate subtraction techniques in detail, nor the residual resulting from the subtraction, but assume that 
the signals can be removed with high enough accuracy to search for an underlying stochastic
gravitational wave background of a different origin.

{\em Simulation of a population} ---
In order to calculate the total contribution of BBHs to the confusion background, 
we consider the fiducial model of Ref.\,\cite{gw150914sgwb} and generate an extra-galactic
population of BBHs using the Monte Carlo procedure described in
\cite{ETMDC1, ETMDC2, 2015PhRvD..92f3002M} and summarized below. 
\begin{itemize}
\item The intrinsic masses $m_1$, $m_2$ (in the source frame) are selected from one of the two astrophysical
distributions considered in Ref.\,\cite{2016arXiv160604856T}: 
%
%
(i) model A: power-law distribution of the primary (i.e., larger mass) 
companion $p(m_1) \propto m_1^{-2.35},$ and uniform distribution of the secondary and (ii) model B: uniform distribution in the logarithm of the component masses $p(m_1,m_2) \propto
m_1^{-1} m_2^{-1}$. In
addition, we require that the component masses take values in the range 5--100\,
M$_\odot$ with $m_1+m_2<100$\, M$_\odot$.

\item The redshift is drawn from a probability distribution $p(z)$
\begin{equation}
p(z)=\frac{R_z(z)}{\int_0^{20} R_z(z) dz}
\end{equation}
obtained by normalizing the merger rate (in the observer
frame) per interval of redshift, over the range $z\in$ 0--20, and
\begin{equation}
R_z(z) = \int \frac{R_m(z)}{1+z} \frac{dV}{dz}(z) dz. 
\end{equation}
Here $\frac{dV}{dz}$ is the comoving volume element and $R_m$ (in the source
frame) is the rate per volume, given by:
\begin{equation}
R_m(z) = \int_{t_{\min}}^{t_{\max}} R_f(z_f)P(t_d) dt_d, 
\end{equation}
where $R_f(z)$ is the massive binary formation rate, $P(t_d)$ the distribution
of the time delay $t_d$  between the formation of the massive progenitors and
their merger, $z_f$ is the redshift at the formation time $t_f=t(z)-t_d$, and $t(z)$
is the age of the Universe at merger. The value of $R_m$ at $z=0$ corresponds to
the local rate estimated from the first LIGO observation run
\cite{2016arXiv160604856T}, which is $99^{+138}_{-70}$ Gpc$^{-3}$yr$^{-1}$ for model A and $30^{+43}_{-21}$ Gpc$^{-3}$yr$^{-1}$ for model B.

We assume that $R_f(z) $ follows the cosmic star formation rate and we use the
recent model of \cite{2015MNRAS.447.2575V}, based on the gamma-ray burst rate of
\cite{kistler} and on the normalization described in \cite{trenti,behroozi}. We
also assume that black holes of 30 M$_\odot$ or larger can only be
formed below the metallicity threshold $Z_c = Z_{\odot}/2$
\cite{gw150914astro,gw150914sgwb}. 
The metallicity is drawn from a $\log_{10}$-normal distribution with a standard
deviation of 0.5 around the mean at each redshift \cite{2015MNRAS.452L..36D}
calculated from the mean metallicity-redshift relation of
Ref.\,\cite{2014ARAA..52..415M}, rescaled upwards by a factor of 3 to account for
local observations \cite{gw150914chris,2015MNRAS.447.2575V}. 
We further assume that the time delay distribution follows
 $P(t_d) \propto t_d^\alpha$,
with $\alpha=-1$ for $t_d>t_{\min}$ \cite{2002ApJ...572..407B,2004JCAP...06..007A,2006ApJ...648.1110B,2006IJMPD..15..235D,2007ApJ...664.1000B,2007PhR...442..166N,2008ApJ...675..566O,2012ApJ...759...52D}, where $t_{\min}=\unit[50]{Myr}$ is the minimum delay time for a massive binary to evolve until coalescence \cite[e.g.,][]{dominik}, and a maximum time delay $t_{\max}$ equal to the Hubble time.  
 
\item The location in the sky $\hat{\Omega}$, the cosine of the orientation
$\iota$, the polarization $\psi$ and the phase of the signal at coalescence $\phi_0,$ were
drawn from uniform distributions.

\item For each BBH, we determine if its resultant GW emission is detectable in a 
given detector network. The
signal-to-noise ratio (SNR), $\rho_A$, detected by matched filtering with an optimum filter
in the ideal case of Gaussian noise, in a detector labelled $A$, is: 
\begin{equation}
\rho_{A}^2 = 4 \int_0^\infty \frac{\left |F_{+,A}\tilde{h}_{+}
 +F_{\times,A}\tilde{h}_{\times} \right |^2}{S_{n,A}}\, {\rm d}f,
\end{equation}
where $f$ is the GW frequency in the observer frame,
$\tilde{h}_{+}$ and $\tilde{h}_{\times}$ the Fourier transforms
of the GW strain amplitudes of + and $\times$ polarisations that
includes inspiral, merger and ringdown phases of the signal 
\cite{2011PhRvD..84h4037A}, $F_{+,A}$ and $F_{\times,A}$ are the
antenna response functions to the GW $+$ and $\times $ polarisations, 
and $S_{n,A}(f)$ is the one-sided noise power spectral density (PSD) 
of detector A. The coherent SNR for
a network, assuming uncorrelated noises in the detectors, is simply given by the
quadrature sum of the individual SNRs $\rho_T^2 = \sum \rho_A^2.$
We assume that sources with $\rho_T>12$  can be removed with enough accuracy from the data and that only sources with $\rho_T<12$ contribute to the confusion background. We are currently investigating this assumption 
using mock data challenges.
\end{itemize}

\begin{figure}
\includegraphics[width=0.45\textwidth]{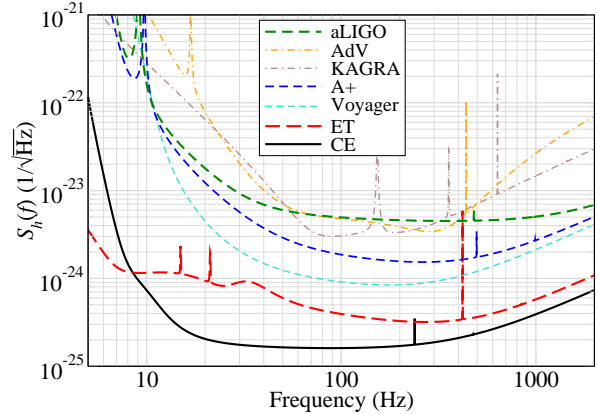}
\caption{Design power spectral density of second generation detectors: Advanced
LIGO (aLIGO), Advanced Virgo (AdV) and KAGRA and proposed sensitivity of
third generation detectors Einstein Telescope (ET) and Cosmic Explorer or (CE).
Expected intermediate sensitivities such as Advanced LIGO Plus (A+) and Voyager 
are also shown.}
\label{fig:sensitivity}
\end{figure}

{\em Detected sources} ---
In this section we investigate the evolution of the number of detections as the detector sensitivity increases from second to third generation and the number of
detectors in the network increases from three to five. 
The Advanced version of the two LIGO detectors at Hanford (H) and
Livingston (L)  \cite{cqg.27.084006.10,cqg.32.074001.15} started collecting
data in September 2015 and are expected to reach design sensitivity in 2019,
followed by Advanced Virgo (V) a few months later \cite{ObsScenario}. Two other
detectors will join the network over the next eight years: the
Japanese detector KAGRA (K) \cite{2013PhRvD..88d3007A} and a new detector in 
India (I)\cite{Indigo} whose sensitivity will be similar to the two LIGO detectors. Third generation
detectors are currently under design study, such as the Einstein
Telescope (ET) \cite{cqg.27.194002.10}, 
and the Cosmic Explorer (CE) \cite{2016arXiv160708697A}.
Between the second and the third generation we expect to reach intermediate
sensitivities, referred to as A+ and Voyager. Figure \ref{fig:sensitivity} plots
the strain sensitivity of the various detectors considered in this paper.

The total number of BBHs that coalesce in the observable Universe, as derived from the actual constraints on the local rate \cite{2016arXiv160604856T}, is in the range  $[1,40]\times 10^4$ a year; therefore, the average waiting time between two consecutive events in the detector frame is between $100$ and $2000$ s. With advanced detectors only a small fraction of the sources will be detected (less than 3\% assuming the maximal rate) and the chance that two detections overlap in time is very small (less than 0.05\%) given that the average duration of the signal is only 0.7 s (model A) or 0.2 s (model B) with the low frequency limit of the detectors of 10 Hz.
With a network of 3rd generation detectors, on the other hand, most of the sources will be above the detection threshold (more than 99.9\%) and the signal will last much longer since the low frequency limit will be pushed to about 5 Hz. For the average rates, we expect 27\% of sources to have some overlap (model A with an average duration of 83 s) or 3.5\% (model B  with an average duration of 26 s) and up to 48\% (model A ) or 11\% (model B) for the maximal rates. However, when there is an overlap in the time domain the sources can still be resolved individually. In fact, the low frequency part of the signal contributes little to the SNR. In the frequency band starting at 20 Hz, we have more than a 99\% of chance of detection while decreasing the chance of overlap to 0.8\% (model A with an average duration of  2 s) or 0.25\% (model B with an average duration of 0.65 s) for the average rate, and about 2.3\% (model A) and 1\% (model B) for the maximal rate.


{\em Binary background} ---
The superposition of the gravitational waves from sources at all redshifts and integrated
over all directions of the sky, creates a stochastic background, whose energy-density spectrum
in GWs is described by the dimensionless quantity
\cite{1999PhRvD..59j2001A}:
\begin{eqnarray}
  \Omega_\text{GW}(f) = \frac{f}{\rho_c} \frac{d\rho_\text{GW}}{df}\,,
\end{eqnarray}
where  $d\rho_\text{GW}$ is the energy density in the frequency
interval $f$ to $f+df$, $\rho _{c} = 3H_0^2c^2/8\pi G$ is the closure energy
density of the Universe, and $H_0=67.8 \pm 0.9{\rm \; km/s/Mpc}$
is the Hubble constant \cite{Ade:2015xua}.

The GW spectrum from the population of BBHs is given by the expression: 
\begin{equation}
\Omega_{\rm{GW}}(f)=\frac{1}{\rho_c c} f F(f).
\label{eq:omega_flux}
\end{equation}
where $F(f)$ is the total flux and $f$ is the observed frequency.
The total flux (in erg Hz$^{-1}$) is the sum of the individual contributions:  
\begin{equation}
F(f)= T^{-1}  \frac{\pi c^3}{2G} f^2 \sum_{k=1}^{N} (\tilde{h}^2_{+,k}(f) +
\tilde{h}^2_{\times,k}(f)	)
\label{eq:flux}
\end{equation}
where $N$ is the number of undetected sources in the Monte Carlo sample (in order to obtain a smooth average of the spectrum we set $N=10^5$ for the sources with $\rho_T < 12)$. The normalization factor $T^{-1}$ assures that the flux has the correct dimension, $T$ being the length of the data sample.

\begin{figure}
\includegraphics[width=0.45\textwidth]{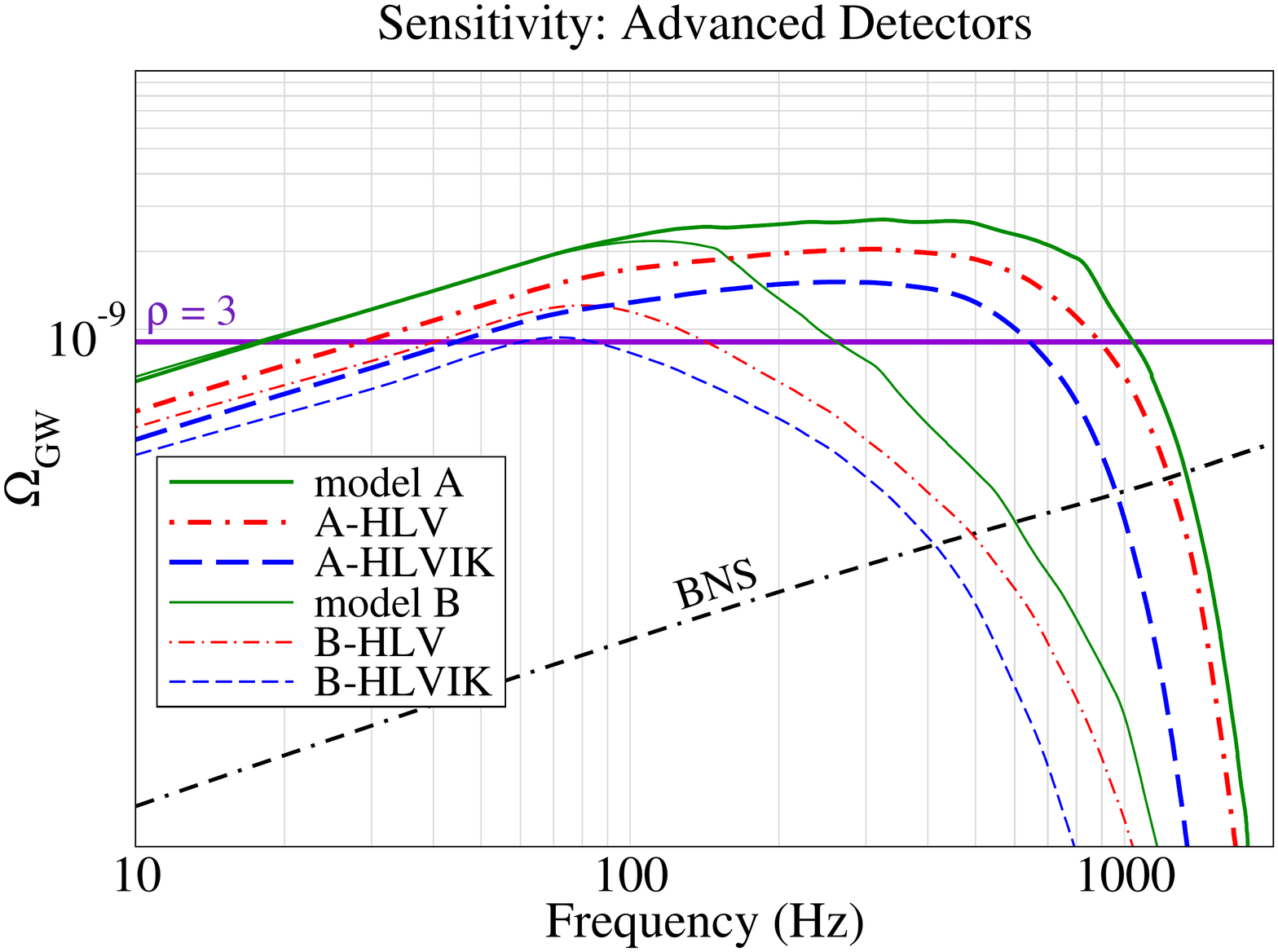}
\includegraphics[width=0.45\textwidth]{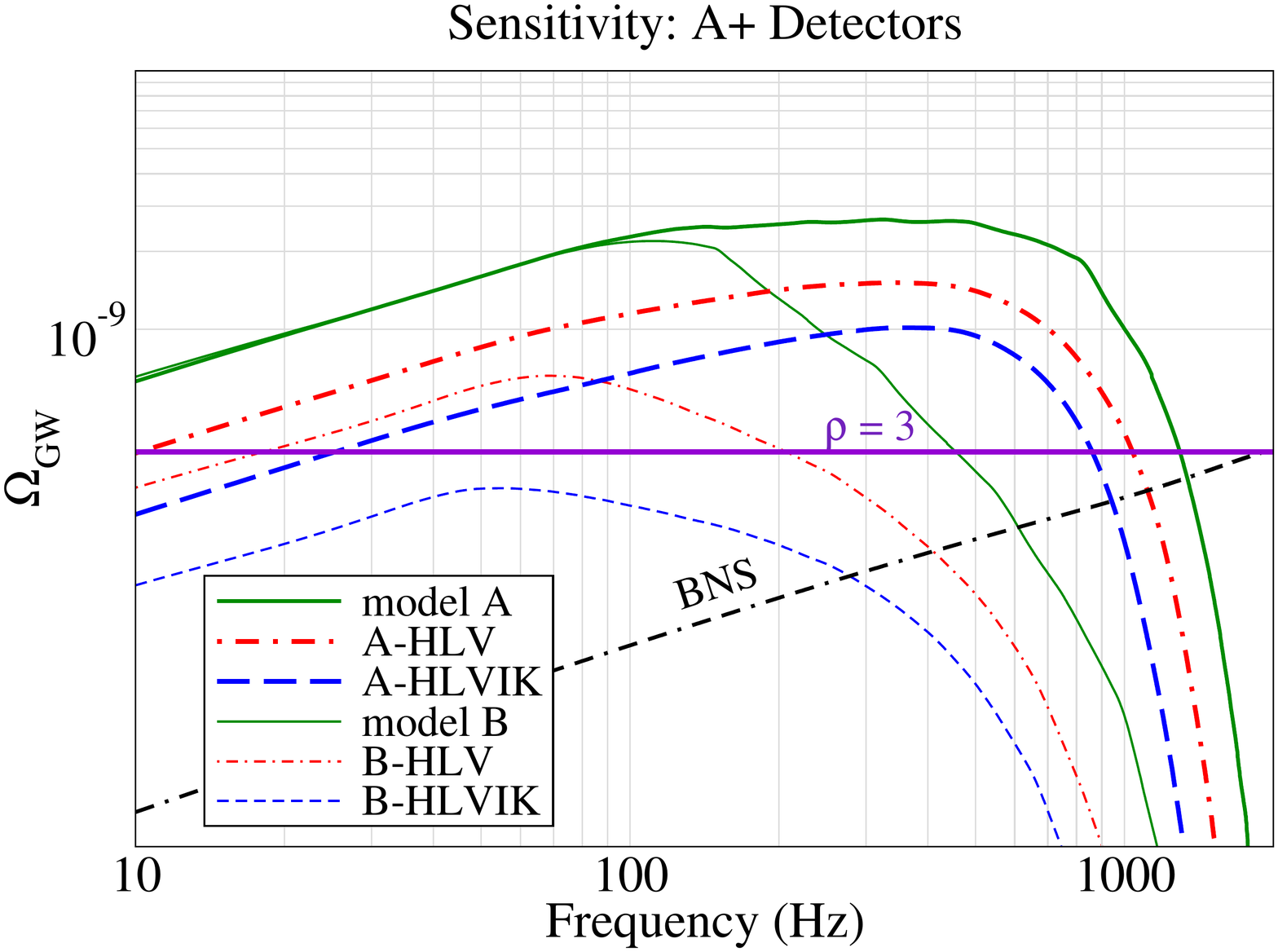}\\
\includegraphics[width=0.45\textwidth]{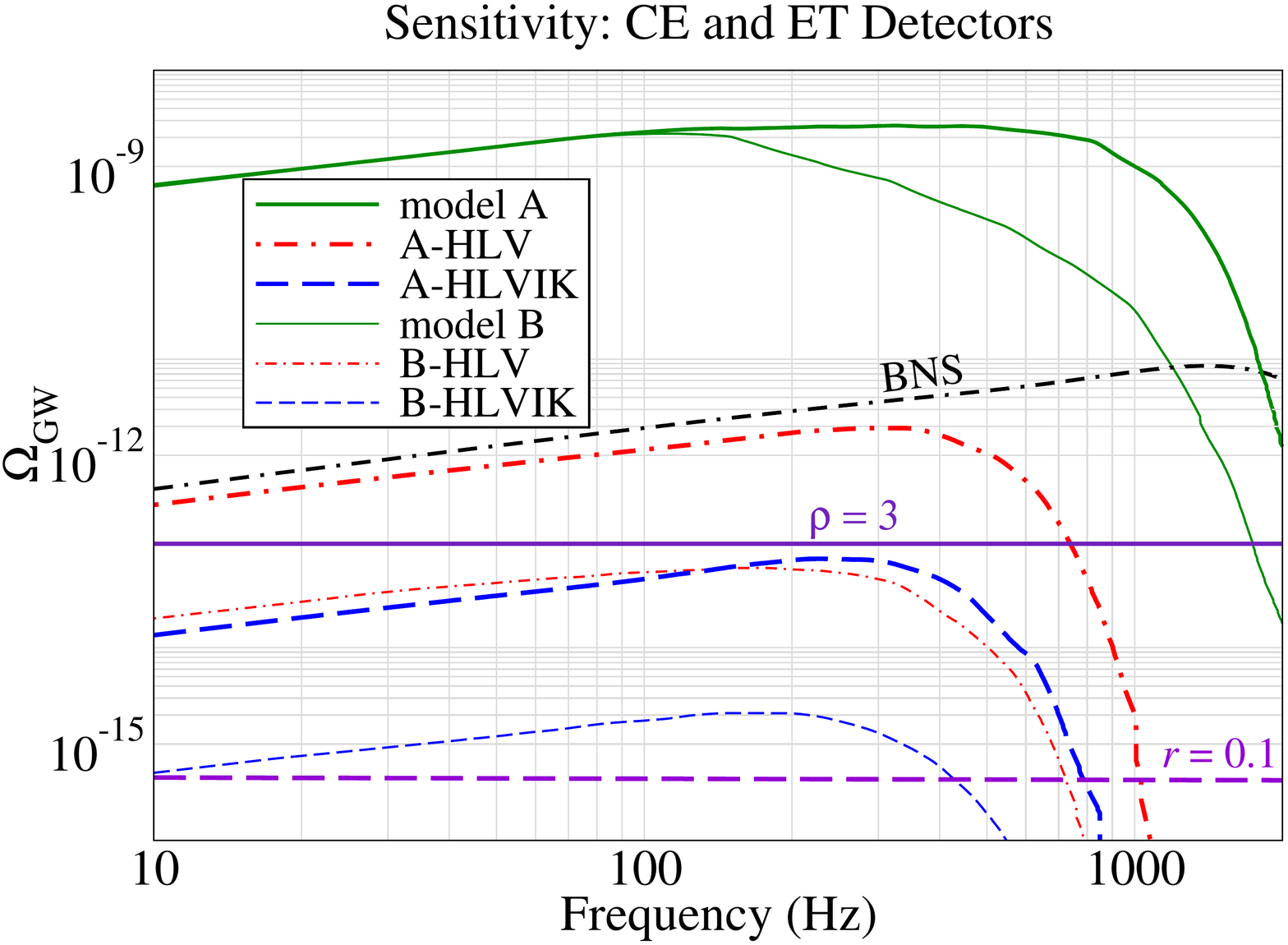}
\caption{Energy density spectrum $\Omega_{\rm GW}$ in GWs from undetected BBHs ($\rho_T<12$) 
with in Advanced (top plot), A+ (middle plot) and third generation (bottom plot) detectors. 
Solid (green) curves are the total backgrounds for models
A (thick lines) and B (thin lines), respectively, when detected BBH signals are not removed from the data,
so they are the same in each plot. We see that in the tens of Hz region one obtains the characteristic $f^{2/3}$ slope.
The cosmological background from inflation assuming a tensor-to-scalar
ratio of $r=0.1$ is shown for comparison, and confusion background
from unresolved binary neutron stars, assuming an average local rate of
60 Gpc$^{-3}$ yr$^{-1}$ \cite{2015ApJ...806..263D}. The horizontal solid 
line is the minimal flat spectrum that can be detected with $\rho =  3$ with a 5-detector network after five years.}
\label{fig:omega} 
\end{figure}

\begin{widetext}
Our waveform model includes inspiral, merger and ringdown phases of the signal. 
In the inspiral regime, before the black holes reach the last stable orbit, the
slope of the spectrum has the well-known $f^{2/3}$ behavior: 
\begin{equation}
\Omega_{\rm{GW}}^{\rm{insp}}(f) = \frac{5 \pi^{2/3} G^{5/3}c^{5/3}}{18 c^3 H_0^2} T^{-1} f^{2/3}
\sum_{k=1}^{N} \frac{(1+z_k)^{5/3} (\mathcal{M}_k)^{5/3}}{D_{\rm
L}(z_k)^2} \left [\frac{(1+\cos^2 \iota_k)^2}{4}+\cos^2\iota_k \right ]
\end{equation}
where $M=m_1+m_2$ is the total mass, $\mathcal{M}=(m_1 m_2)^{3/5}M^{-1/5}$ the
chirp mass and $D_L(z)$ is the luminosity distance at redshift $z.$ We shall
see below that we retrieve this behavior over the relevant range of frequencies.
\end{widetext}
Fig.~\ref{fig:omega} shows the energy density $\Omega_{\rm GW}$ in GWs 
from undetected BBHs ($\rho_T<12$) with in Advanced (top plot), A+ (middle plot) and third generation (bottom plot) detectors. 
Solid (green) curves are the total backgrounds for models
A (thick lines) and B (thin lines), respectively, when detected BBH signals are not removed from the data,
so they are the same in each plot. 
For each generation of sensitivity, we consider two different networks: A network of 
3 detectors (HLV) located at the sites of LIGO-Hanford, LIGO-Livingston and Virgo
and a network of 5 detectors (HLVIK) that includes LIGO India and KAGRA, in 
addition to HLV. In the top plot, the detectors are assumed to have 
projected sensitivity levels of advanced detectors shown in Fig.~\ref{fig:sensitivity}.  
In the middle plot, we assume that all the detectors
have the same intermediate sensitivity (A+). In the bottom plot, for the third generation 
we assume the sensitivity of ET in a triangle detector configuration at the location of Virgo and CE for all other detectors.

{\em Results} ---
The total background from BBHs when detected BBH signals are not removed from the data is expected to dominate over all other sources of stochastic background, up to a few hundred of Hz with an average energy density of $\Omega_{\rm GW}(\mathrm{10\,Hz})  = 6 \times 10^{-10}$. With Advanced detectors the BBH confusion background is more than 50\% of the total background, and still above $\Omega_{\rm GW}(\mathrm{ 10\,Hz}) = 10^{-10}$ with the A+ sensitivity. 
With third generation detectors, on the other hand, the level of the confusion background is decreased by orders of magnitude, reaching $\Omega_{\rm GW}(\mathrm{10\,Hz})  = 10^{-14}-10^{-13}$ with a network of 3 detectors and $\Omega_{\rm GW}(\mathrm{10\,Hz}) =10^{-16}-10^{-14}$ with 5 detectors.

With a network of 5 third generation detectors we are able to decrease the confusion background below the minimal detectable flat energy spectrum (5 years of integration) of $\Omega_{\rm GW}({\mathrm{ 10 Hz}})  = 10^{-13}$.
The detectable value is derived requiring a SNR of $\rho=3$ where 
\cite{gw150914sgwb}
\begin{equation}
\rho =\frac{3 H_0^2}{10 \pi^2} \sqrt{2T} \left[
\int_0^\infty df\>
\sum_{i=1}^n\sum_{j>i}
\frac{\gamma_{ij}^2(f)\Omega_{\rm GW}^2(f)}{f^6 S_{n,i}(f)S_{n,j}(f)} \right]^{1/2}\,,
\label{eq:snrCC}
\end{equation}
for a network of detectors $i=1,2,\cdots, n$. In this expression $\gamma_{ij}$ is the overlap reduction function characterizing the reduction of sensitivity due to the separation and the relative orientation of the detectors. The contribution to the SNR comes mostly from the closest pair of detectors, namely the LIGO Hanford -- LIGO Livingston detector pairs over a frequency interval of 50 Hz. The entire network of detectors is needed to identify the signals, estimate their parameters~\cite{2016arXiv161006917V}, and then remove their presence from the data.
This minimal detectable value is above the current upper limit for the standard inflation model assuming a tensor-to-scalar ratio $r=0.1$, meaning that the detectors' sensitivity should be improved by at least another factor of about 10  to reach a level of $\Omega_{\min} \sim 10^{-15}$.

An improvement by a factor of 10 in sensitivity past ET and CE  would also allow for the removal of the extra confusion background from BNSs that could remain in the data at the level of $\Omega_{\mathrm{GW}}(\rm{10 Hz}) = 4.5 \times 10^{-13} $, as shown in Fig.\ref{fig:omega}.
The level of this confusion background is uncertain but future detections will provide constraints on the rate of such events and allow for more accurate predictions.

{\em Conclusions } ---
In this study we have demonstrated that third generation GW
detectors will have sensitivities sufficient to directly observe almost every
coalescing BBH system in the Universe. However a more detailed analysis is needed to assess how well one can subtract BBH signals from the data, for example using methods similar to those developed for the Big Bang Observer \cite{2006PhRvD..73d2001C,2008PhRvD..77l3010H} or LISA~\cite{PhysRevD.72.022001}; this will be addressed in an ongoing  mock data challenge. With the binary black hole coalescences removed,
these detectors would be sensitive to a PGWB at the level of $\Omega_{\mathrm{GW}} \simeq 10^{-13}$, after five years of observation, comparable to the sensitivity of LISA \cite{2012JCAP...06..027B}. A potential limitation to this sensitivity comes from other astrophysically produced GW, such as those from the coalescence of binary neutron stars, but there is still much uncertainty on the
magnitude of this background. Observations of compact binary coalescence events
in the coming years will provide the necessary information on their merger rate.
The removal of BBH confusion background with third
generation detectors opens up the possibility to observe the PGWB.

{\em Acknowledgments}  ---
We thank Thomas Dent, Vuk Mandic and Alan Weinstein for comments. B.S.S acknowledges the support of Science and Technologies Facilities Council grant ST/L000962/1. N.C received support from NSF grant PHY-1505373. M.E., E.K. and S.V. acknowledge the support of the National Science Foundation and the LIGO Laboratory. LIGO was constructed by the California Institute of Technology and Massachusetts Institute of Technology with funding from the National Science Foundation and operates under cooperative agreement PHY-0757058T.R acknowledges the LIGO Visitors Program and is grateful to X.O. for useful discussions. 
This article has been assigned LIGO Document number P1600323.

\bibliography{regimbau}

\begin{thebibliography}{66}%
\makeatletter
\providecommand \@ifxundefined [1]{%
 \@ifx{#1\undefined}
}%
\providecommand \@ifnum [1]{%
 \ifnum #1\expandafter \@firstoftwo
 \else \expandafter \@secondoftwo
 \fi
}%
\providecommand \@ifx [1]{%
 \ifx #1\expandafter \@firstoftwo
 \else \expandafter \@secondoftwo
 \fi
}%
\providecommand \natexlab [1]{#1}%
\providecommand \enquote  [1]{``#1''}%
\providecommand \bibnamefont  [1]{#1}%
\providecommand \bibfnamefont [1]{#1}%
\providecommand \citenamefont [1]{#1}%
\providecommand \href@noop [0]{\@secondoftwo}%
\providecommand \href [0]{\begingroup \@sanitize@url \@href}%
\providecommand \@href[1]{\@@startlink{#1}\@@href}%
\providecommand \@@href[1]{\endgroup#1\@@endlink}%
\providecommand \@sanitize@url [0]{\catcode `\\12\catcode `\$12\catcode
  `\&12\catcode `\#12\catcode `\^12\catcode `\_12\catcode `\%12\relax}%
\providecommand \@@startlink[1]{}%
\providecommand \@@endlink[0]{}%
\providecommand \url  [0]{\begingroup\@sanitize@url \@url }%
\providecommand \@url [1]{\endgroup\@href {#1}{\urlprefix }}%
\providecommand \urlprefix  [0]{URL }%
\providecommand \Eprint [0]{\href }%
\providecommand \doibase [0]{http://dx.doi.org/}%
\providecommand \selectlanguage [0]{\@gobble}%
\providecommand \bibinfo  [0]{\@secondoftwo}%
\providecommand \bibfield  [0]{\@secondoftwo}%
\providecommand \translation [1]{[#1]}%
\providecommand \BibitemOpen [0]{}%
\providecommand \bibitemStop [0]{}%
\providecommand \bibitemNoStop [0]{.\EOS\space}%
\providecommand \EOS [0]{\spacefactor3000\relax}%
\providecommand \BibitemShut  [1]{\csname bibitem#1\endcsname}%
\let\auto@bib@innerbib\@empty
\bibitem [{\citenamefont {{Grishchuk}}(1975)}]{1975JETP...40..409G}%
  \BibitemOpen
  \bibfield  {author} {\bibinfo {author} {\bibfnamefont {L.~P.}\ \bibnamefont
  {{Grishchuk}}},\ }\href@noop {} {\bibfield  {journal} {\bibinfo  {journal}
  {Soviet Journal of Experimental and Theoretical Physics}\ }\textbf {\bibinfo
  {volume} {40}},\ \bibinfo {pages} {409} (\bibinfo {year} {1975})}\BibitemShut
  {NoStop}%
\bibitem [{\citenamefont {{Grishchuk}}(1993)}]{1993PhRvD..48.3513G}%
  \BibitemOpen
  \bibfield  {author} {\bibinfo {author} {\bibfnamefont {L.~P.}\ \bibnamefont
  {{Grishchuk}}},\ }\href {\doibase 10.1103/PhysRevD.48.3513} {\bibfield
  {journal} {\bibinfo  {journal} {\prd}\ }\textbf {\bibinfo {volume} {48}},\
  \bibinfo {pages} {3513} (\bibinfo {year} {1993})},\ \Eprint
  {http://arxiv.org/abs/gr-qc/9304018} {gr-qc/9304018} \BibitemShut {NoStop}%
\bibitem [{\citenamefont {{Starobinskii}}(1979)}]{1979ZhPmR..30..719S}%
  \BibitemOpen
  \bibfield  {author} {\bibinfo {author} {\bibfnamefont {A.~A.}\ \bibnamefont
  {{Starobinskii}}},\ }\href@noop {} {\bibfield  {journal} {\bibinfo  {journal}
  {ZhETF Pisma Redaktsiiu}\ }\textbf {\bibinfo {volume} {30}},\ \bibinfo
  {pages} {719} (\bibinfo {year} {1979})}\BibitemShut {NoStop}%
\bibitem [{\citenamefont {{Gasperini}}\ and\ \citenamefont
  {{Veneziano}}(1993)}]{1993APh.....1..317G}%
  \BibitemOpen
  \bibfield  {author} {\bibinfo {author} {\bibfnamefont {M.}~\bibnamefont
  {{Gasperini}}}\ and\ \bibinfo {author} {\bibfnamefont {G.}~\bibnamefont
  {{Veneziano}}},\ }\href {\doibase 10.1016/0927-6505(93)90017-8} {\bibfield
  {journal} {\bibinfo  {journal} {Astroparticle Physics}\ }\textbf {\bibinfo
  {volume} {1}},\ \bibinfo {pages} {317} (\bibinfo {year} {1993})},\ \Eprint
  {http://arxiv.org/abs/hep-th/9211021} {hep-th/9211021} \BibitemShut {NoStop}%
\bibitem [{\citenamefont {{Buonanno}}\ \emph {et~al.}(1997)\citenamefont
  {{Buonanno}}, \citenamefont {{Maggiore}},\ and\ \citenamefont
  {{Ungarelli}}}]{1997PhRvD..55.3330B}%
  \BibitemOpen
  \bibfield  {author} {\bibinfo {author} {\bibfnamefont {A.}~\bibnamefont
  {{Buonanno}}}, \bibinfo {author} {\bibfnamefont {M.}~\bibnamefont
  {{Maggiore}}}, \ and\ \bibinfo {author} {\bibfnamefont {C.}~\bibnamefont
  {{Ungarelli}}},\ }\href {\doibase 10.1103/PhysRevD.55.3330} {\bibfield
  {journal} {\bibinfo  {journal} {\prd}\ }\textbf {\bibinfo {volume} {55}},\
  \bibinfo {pages} {3330} (\bibinfo {year} {1997})},\ \Eprint
  {http://arxiv.org/abs/gr-qc/9605072} {gr-qc/9605072} \BibitemShut {NoStop}%
\bibitem [{\citenamefont {{Dufaux}}\ \emph {et~al.}(2010)\citenamefont
  {{Dufaux}}, \citenamefont {{Figueroa}},\ and\ \citenamefont
  {{Garc{\'{\i}}a-Bellido}}}]{2010PhRvD..82h3518D}%
  \BibitemOpen
  \bibfield  {author} {\bibinfo {author} {\bibfnamefont {J.-F.}\ \bibnamefont
  {{Dufaux}}}, \bibinfo {author} {\bibfnamefont {D.~G.}\ \bibnamefont
  {{Figueroa}}}, \ and\ \bibinfo {author} {\bibfnamefont {J.}~\bibnamefont
  {{Garc{\'{\i}}a-Bellido}}},\ }\href {\doibase 10.1103/PhysRevD.82.083518}
  {\bibfield  {journal} {\bibinfo  {journal} {\prd}\ }\textbf {\bibinfo
  {volume} {82}},\ \bibinfo {eid} {083518} (\bibinfo {year} {2010})},\ \Eprint
  {http://arxiv.org/abs/1006.0217} {arXiv:1006.0217 [astro-ph.CO]} \BibitemShut
  {NoStop}%
\bibitem [{\citenamefont {{Damour}}\ and\ \citenamefont
  {{Vilenkin}}(2005)}]{2005PhRvD..71f3510D}%
  \BibitemOpen
  \bibfield  {author} {\bibinfo {author} {\bibfnamefont {T.}~\bibnamefont
  {{Damour}}}\ and\ \bibinfo {author} {\bibfnamefont {A.}~\bibnamefont
  {{Vilenkin}}},\ }\href {\doibase 10.1103/PhysRevD.71.063510} {\bibfield
  {journal} {\bibinfo  {journal} {\prd}\ }\textbf {\bibinfo {volume} {71}},\
  \bibinfo {eid} {063510} (\bibinfo {year} {2005})},\ \Eprint
  {http://arxiv.org/abs/hep-th/0410222} {hep-th/0410222} \BibitemShut {NoStop}%
\bibitem [{\citenamefont {{Siemens}}\ \emph {et~al.}(2007)\citenamefont
  {{Siemens}}, \citenamefont {{Mandic}},\ and\ \citenamefont
  {{Creighton}}}]{2007PhRvL..98k1101S}%
  \BibitemOpen
  \bibfield  {author} {\bibinfo {author} {\bibfnamefont {X.}~\bibnamefont
  {{Siemens}}}, \bibinfo {author} {\bibfnamefont {V.}~\bibnamefont {{Mandic}}},
  \ and\ \bibinfo {author} {\bibfnamefont {J.}~\bibnamefont {{Creighton}}},\
  }\href {\doibase 10.1103/PhysRevLett.98.111101} {\bibfield  {journal}
  {\bibinfo  {journal} {Physical Review Letters}\ }\textbf {\bibinfo {volume}
  {98}},\ \bibinfo {eid} {111101} (\bibinfo {year} {2007})},\ \Eprint
  {http://arxiv.org/abs/astro-ph/0610920} {astro-ph/0610920} \BibitemShut
  {NoStop}%
\bibitem [{\citenamefont {{{\"O}lmez}}\ \emph {et~al.}(2010)\citenamefont
  {{{\"O}lmez}}, \citenamefont {{Mandic}},\ and\ \citenamefont
  {{Siemens}}}]{2010PhRvD..81j4028O}%
  \BibitemOpen
  \bibfield  {author} {\bibinfo {author} {\bibfnamefont {S.}~\bibnamefont
  {{{\"O}lmez}}}, \bibinfo {author} {\bibfnamefont {V.}~\bibnamefont
  {{Mandic}}}, \ and\ \bibinfo {author} {\bibfnamefont {X.}~\bibnamefont
  {{Siemens}}},\ }\href {\doibase 10.1103/PhysRevD.81.104028} {\bibfield
  {journal} {\bibinfo  {journal} {\prd}\ }\textbf {\bibinfo {volume} {81}},\
  \bibinfo {eid} {104028} (\bibinfo {year} {2010})},\ \Eprint
  {http://arxiv.org/abs/1004.0890} {arXiv:1004.0890 [astro-ph.CO]} \BibitemShut
  {NoStop}%
\bibitem [{\citenamefont {{Regimbau}}\ \emph
  {et~al.}(2012{\natexlab{a}})\citenamefont {{Regimbau}}, \citenamefont
  {{Giampanis}}, \citenamefont {{Siemens}},\ and\ \citenamefont
  {{Mandic}}}]{2012PhRvD..85f6001R}%
  \BibitemOpen
  \bibfield  {author} {\bibinfo {author} {\bibfnamefont {T.}~\bibnamefont
  {{Regimbau}}}, \bibinfo {author} {\bibfnamefont {S.}~\bibnamefont
  {{Giampanis}}}, \bibinfo {author} {\bibfnamefont {X.}~\bibnamefont
  {{Siemens}}}, \ and\ \bibinfo {author} {\bibfnamefont {V.}~\bibnamefont
  {{Mandic}}},\ }\href {\doibase 10.1103/PhysRevD.85.066001} {\bibfield
  {journal} {\bibinfo  {journal} {\prd}\ }\textbf {\bibinfo {volume} {85}},\
  \bibinfo {eid} {066001} (\bibinfo {year} {2012}{\natexlab{a}})},\ \Eprint
  {http://arxiv.org/abs/1111.6638} {arXiv:1111.6638 [astro-ph.CO]} \BibitemShut
  {NoStop}%
\bibitem [{\citenamefont {{Caprini}}\ \emph {et~al.}(2008)\citenamefont
  {{Caprini}}, \citenamefont {{Durrer}},\ and\ \citenamefont
  {{Servant}}}]{2008PhRvD..77l4015C}%
  \BibitemOpen
  \bibfield  {author} {\bibinfo {author} {\bibfnamefont {C.}~\bibnamefont
  {{Caprini}}}, \bibinfo {author} {\bibfnamefont {R.}~\bibnamefont {{Durrer}}},
  \ and\ \bibinfo {author} {\bibfnamefont {G.}~\bibnamefont {{Servant}}},\
  }\href {\doibase 10.1103/PhysRevD.77.124015} {\bibfield  {journal} {\bibinfo
  {journal} {\prd}\ }\textbf {\bibinfo {volume} {77}},\ \bibinfo {eid} {124015}
  (\bibinfo {year} {2008})},\ \Eprint {http://arxiv.org/abs/0711.2593}
  {arXiv:0711.2593} \BibitemShut {NoStop}%
\bibitem [{\citenamefont {{Caprini}}\ \emph
  {et~al.}(2009{\natexlab{a}})\citenamefont {{Caprini}}, \citenamefont
  {{Durrer}}, \citenamefont {{Konstandin}},\ and\ \citenamefont
  {{Servant}}}]{2009PhRvD..79h3519C}%
  \BibitemOpen
  \bibfield  {author} {\bibinfo {author} {\bibfnamefont {C.}~\bibnamefont
  {{Caprini}}}, \bibinfo {author} {\bibfnamefont {R.}~\bibnamefont {{Durrer}}},
  \bibinfo {author} {\bibfnamefont {T.}~\bibnamefont {{Konstandin}}}, \ and\
  \bibinfo {author} {\bibfnamefont {G.}~\bibnamefont {{Servant}}},\ }\href
  {\doibase 10.1103/PhysRevD.79.083519} {\bibfield  {journal} {\bibinfo
  {journal} {\prd}\ }\textbf {\bibinfo {volume} {79}},\ \bibinfo {eid} {083519}
  (\bibinfo {year} {2009}{\natexlab{a}})},\ \Eprint
  {http://arxiv.org/abs/0901.1661} {arXiv:0901.1661 [astro-ph.CO]} \BibitemShut
  {NoStop}%
\bibitem [{\citenamefont {{Caprini}}\ \emph
  {et~al.}(2009{\natexlab{b}})\citenamefont {{Caprini}}, \citenamefont
  {{Durrer}},\ and\ \citenamefont {{Servant}}}]{2009JCAP...12..024C}%
  \BibitemOpen
  \bibfield  {author} {\bibinfo {author} {\bibfnamefont {C.}~\bibnamefont
  {{Caprini}}}, \bibinfo {author} {\bibfnamefont {R.}~\bibnamefont {{Durrer}}},
  \ and\ \bibinfo {author} {\bibfnamefont {G.}~\bibnamefont {{Servant}}},\
  }\href {\doibase 10.1088/1475-7516/2009/12/024} {\bibfield  {journal}
  {\bibinfo  {journal} {Journal of Cosmology and Astroparticle Physics}\
  }\textbf {\bibinfo {volume} {12}},\ \bibinfo {eid} {024} (\bibinfo {year}
  {2009}{\natexlab{b}})},\ \Eprint {http://arxiv.org/abs/0909.0622}
  {arXiv:0909.0622 [astro-ph.CO]} \BibitemShut {NoStop}%
\bibitem [{\citenamefont {{Regimbau}}(2011)}]{2011RAA....11..369R}%
  \BibitemOpen
  \bibfield  {author} {\bibinfo {author} {\bibfnamefont {T.}~\bibnamefont
  {{Regimbau}}},\ }\href {\doibase 10.1088/1674-4527/11/4/001} {\bibfield
  {journal} {\bibinfo  {journal} {Research in Astronomy and Astrophysics}\
  }\textbf {\bibinfo {volume} {11}},\ \bibinfo {pages} {369} (\bibinfo {year}
  {2011})},\ \Eprint {http://arxiv.org/abs/1101.2762} {arXiv:1101.2762
  [astro-ph.CO]} \BibitemShut {NoStop}%
\bibitem [{\citenamefont {Punturo}\ \emph {et~al.}(2010)\citenamefont {Punturo}
  \emph {et~al.}}]{cqg.27.194002.10}%
  \BibitemOpen
  \bibfield  {author} {\bibinfo {author} {\bibfnamefont {M.}~\bibnamefont
  {Punturo}} \emph {et~al.},\ }\href {\doibase 10.1088/0264-9381/27/19/194002}
  {\bibfield  {journal} {\bibinfo  {journal} {Classical Quantum Gravity}\
  }\textbf {\bibinfo {volume} {27}},\ \bibinfo {pages} {194002} (\bibinfo
  {year} {2010})}\BibitemShut {NoStop}%
\bibitem [{\citenamefont {{Abbott}}\ \emph
  {et~al.}(2016{\natexlab{a}})\citenamefont {{Abbott}}, \citenamefont
  {{Abbott}}, \citenamefont {{Abbott}}, \citenamefont {{Abernathy}},
  \citenamefont {{Ackley}}, \citenamefont {{Adams}}, \citenamefont {{Addesso}},
  \citenamefont {{Adhikari}}, \citenamefont {{Adya}}, \citenamefont
  {{Affeldt}},\ and\ \citenamefont {et~al.}}]{2016arXiv160708697A}%
  \BibitemOpen
  \bibfield  {author} {\bibinfo {author} {\bibfnamefont {B.~P.}\ \bibnamefont
  {{Abbott}}}, \bibinfo {author} {\bibfnamefont {R.}~\bibnamefont {{Abbott}}},
  \bibinfo {author} {\bibfnamefont {T.~D.}\ \bibnamefont {{Abbott}}}, \bibinfo
  {author} {\bibfnamefont {M.~R.}\ \bibnamefont {{Abernathy}}}, \bibinfo
  {author} {\bibfnamefont {K.}~\bibnamefont {{Ackley}}}, \bibinfo {author}
  {\bibfnamefont {C.}~\bibnamefont {{Adams}}}, \bibinfo {author} {\bibfnamefont
  {P.}~\bibnamefont {{Addesso}}}, \bibinfo {author} {\bibfnamefont {R.~X.}\
  \bibnamefont {{Adhikari}}}, \bibinfo {author} {\bibfnamefont {V.~B.}\
  \bibnamefont {{Adya}}}, \bibinfo {author} {\bibfnamefont {C.}~\bibnamefont
  {{Affeldt}}}, \ and\ \bibinfo {author} {\bibnamefont {et~al.}},\ }\href@noop
  {} {\bibfield  {journal} {\bibinfo  {journal} {ArXiv e-prints}\ } (\bibinfo
  {year} {2016}{\natexlab{a}})},\ \Eprint {http://arxiv.org/abs/1607.08697}
  {arXiv:1607.08697 [astro-ph.IM]} \BibitemShut {NoStop}%
\bibitem [{\citenamefont {{Cutler}}\ and\ \citenamefont
  {{Harms}}(2006)}]{2006PhRvD..73d2001C}%
  \BibitemOpen
  \bibfield  {author} {\bibinfo {author} {\bibfnamefont {C.}~\bibnamefont
  {{Cutler}}}\ and\ \bibinfo {author} {\bibfnamefont {J.}~\bibnamefont
  {{Harms}}},\ }\href {\doibase 10.1103/PhysRevD.73.042001} {\bibfield
  {journal} {\bibinfo  {journal} {\prd}\ }\textbf {\bibinfo {volume} {73}},\
  \bibinfo {eid} {042001} (\bibinfo {year} {2006})},\ \Eprint
  {http://arxiv.org/abs/gr-qc/0511092} {gr-qc/0511092} \BibitemShut {NoStop}%
\bibitem [{\citenamefont {{Harms}}\ \emph {et~al.}(2008)\citenamefont
  {{Harms}}, \citenamefont {{Mahrdt}}, \citenamefont {{Otto}},\ and\
  \citenamefont {{Prie{\ss}}}}]{2008PhRvD..77l3010H}%
  \BibitemOpen
  \bibfield  {author} {\bibinfo {author} {\bibfnamefont {J.}~\bibnamefont
  {{Harms}}}, \bibinfo {author} {\bibfnamefont {C.}~\bibnamefont {{Mahrdt}}},
  \bibinfo {author} {\bibfnamefont {M.}~\bibnamefont {{Otto}}}, \ and\ \bibinfo
  {author} {\bibfnamefont {M.}~\bibnamefont {{Prie{\ss}}}},\ }\href {\doibase
  10.1103/PhysRevD.77.123010} {\bibfield  {journal} {\bibinfo  {journal}
  {\prd}\ }\textbf {\bibinfo {volume} {77}},\ \bibinfo {eid} {123010} (\bibinfo
  {year} {2008})},\ \Eprint {http://arxiv.org/abs/0803.0226} {arXiv:0803.0226
  [gr-qc]} \BibitemShut {NoStop}%
\bibitem [{\citenamefont {{Abbott}}\ \emph
  {et~al.}(2016{\natexlab{b}})\citenamefont {{Abbott}}, \citenamefont
  {{Abbott}}, \citenamefont {{Abbott}}, \citenamefont {{Abernathy}},
  \citenamefont {{Acernese}}, \citenamefont {{Ackley}}, \citenamefont
  {{Adams}}, \citenamefont {{Adams}}, \citenamefont {{Addesso}}, \citenamefont
  {{Adhikari}},\ and\ \citenamefont {et~al.}}]{2016PhRvL.116m1103A}%
  \BibitemOpen
  \bibfield  {author} {\bibinfo {author} {\bibfnamefont {B.~P.}\ \bibnamefont
  {{Abbott}}}, \bibinfo {author} {\bibfnamefont {R.}~\bibnamefont {{Abbott}}},
  \bibinfo {author} {\bibfnamefont {T.~D.}\ \bibnamefont {{Abbott}}}, \bibinfo
  {author} {\bibfnamefont {M.~R.}\ \bibnamefont {{Abernathy}}}, \bibinfo
  {author} {\bibfnamefont {F.}~\bibnamefont {{Acernese}}}, \bibinfo {author}
  {\bibfnamefont {K.}~\bibnamefont {{Ackley}}}, \bibinfo {author}
  {\bibfnamefont {C.}~\bibnamefont {{Adams}}}, \bibinfo {author} {\bibfnamefont
  {T.}~\bibnamefont {{Adams}}}, \bibinfo {author} {\bibfnamefont
  {P.}~\bibnamefont {{Addesso}}}, \bibinfo {author} {\bibfnamefont {R.~X.}\
  \bibnamefont {{Adhikari}}}, \ and\ \bibinfo {author} {\bibnamefont
  {et~al.}},\ }\href {\doibase 10.1103/PhysRevLett.116.131103} {\bibfield
  {journal} {\bibinfo  {journal} {Physical Review Letters}\ }\textbf {\bibinfo
  {volume} {116}},\ \bibinfo {eid} {131103} (\bibinfo {year}
  {2016}{\natexlab{b}})},\ \Eprint {http://arxiv.org/abs/1602.03838}
  {arXiv:1602.03838 [gr-qc]} \BibitemShut {NoStop}%
\bibitem [{\citenamefont {Harry}\ \emph {et~al.}(2010)\citenamefont {Harry}
  \emph {et~al.}}]{cqg.27.084006.10}%
  \BibitemOpen
  \bibfield  {author} {\bibinfo {author} {\bibfnamefont {G.~M.}\ \bibnamefont
  {Harry}} \emph {et~al.} (\bibinfo {collaboration} {LIGO Scientific
  Collaboration}),\ }\href {\doibase 10.1088/0264-9381/27/8/084006} {\bibfield
  {journal} {\bibinfo  {journal} {Classical Quantum Gravity}\ }\textbf
  {\bibinfo {volume} {27}},\ \bibinfo {pages} {084006} (\bibinfo {year}
  {2010})}\BibitemShut {NoStop}%
\bibitem [{\citenamefont {Aasi}\ \emph {et~al.}(2015)\citenamefont {Aasi} \emph
  {et~al.}}]{cqg.32.074001.15}%
  \BibitemOpen
  \bibfield  {author} {\bibinfo {author} {\bibfnamefont {J.}~\bibnamefont
  {Aasi}} \emph {et~al.} (\bibinfo {collaboration} {LIGO Scientific
  Collaboration}),\ }\href {\doibase 10.1088/0264-9381/32/7/074001} {\bibfield
  {journal} {\bibinfo  {journal} {Classical Quantum Gravity}\ }\textbf
  {\bibinfo {volume} {32}},\ \bibinfo {pages} {074001} (\bibinfo {year}
  {2015})}\BibitemShut {NoStop}%
\bibitem [{\citenamefont {{Abbott}}\ \emph
  {et~al.}(2016{\natexlab{c}})\citenamefont {{Abbott}}, \citenamefont
  {{Abbott}}, \citenamefont {{Abbott}}, \citenamefont {{Abernathy}},
  \citenamefont {{Acernese}}, \citenamefont {{Ackley}}, \citenamefont
  {{Adams}}, \citenamefont {{Adams}}, \citenamefont {{Addesso}}, \citenamefont
  {{Adhikari}},\ and\ \citenamefont {et~al.}}]{gw150914}%
  \BibitemOpen
  \bibfield  {author} {\bibinfo {author} {\bibfnamefont {B.~P.}\ \bibnamefont
  {{Abbott}}}, \bibinfo {author} {\bibfnamefont {R.}~\bibnamefont {{Abbott}}},
  \bibinfo {author} {\bibfnamefont {T.~D.}\ \bibnamefont {{Abbott}}}, \bibinfo
  {author} {\bibfnamefont {M.~R.}\ \bibnamefont {{Abernathy}}}, \bibinfo
  {author} {\bibfnamefont {F.}~\bibnamefont {{Acernese}}}, \bibinfo {author}
  {\bibfnamefont {K.}~\bibnamefont {{Ackley}}}, \bibinfo {author}
  {\bibfnamefont {C.}~\bibnamefont {{Adams}}}, \bibinfo {author} {\bibfnamefont
  {T.}~\bibnamefont {{Adams}}}, \bibinfo {author} {\bibfnamefont
  {P.}~\bibnamefont {{Addesso}}}, \bibinfo {author} {\bibfnamefont {R.~X.}\
  \bibnamefont {{Adhikari}}}, \ and\ \bibinfo {author} {\bibnamefont
  {et~al.}},\ }\href {\doibase 10.1103/PhysRevLett.116.061102} {\bibfield
  {journal} {\bibinfo  {journal} {Physical Review Letters}\ }\textbf {\bibinfo
  {volume} {116}},\ \bibinfo {eid} {061102} (\bibinfo {year}
  {2016}{\natexlab{c}})},\ \Eprint {http://arxiv.org/abs/1602.03837}
  {arXiv:1602.03837 [gr-qc]} \BibitemShut {NoStop}%
\bibitem [{\citenamefont {{Abbott}}\ \emph
  {et~al.}(2016{\natexlab{d}})\citenamefont {{Abbott}}, \citenamefont
  {{Abbott}}, \citenamefont {{Abbott}}, \citenamefont {{Abernathy}},
  \citenamefont {{Acernese}}, \citenamefont {{Ackley}}, \citenamefont
  {{Adams}}, \citenamefont {{Adams}}, \citenamefont {{Addesso}}, \citenamefont
  {{Adhikari}},\ and\ \citenamefont {et~al.}}]{gw150914PE}%
  \BibitemOpen
  \bibfield  {author} {\bibinfo {author} {\bibfnamefont {B.~P.}\ \bibnamefont
  {{Abbott}}}, \bibinfo {author} {\bibfnamefont {R.}~\bibnamefont {{Abbott}}},
  \bibinfo {author} {\bibfnamefont {T.~D.}\ \bibnamefont {{Abbott}}}, \bibinfo
  {author} {\bibfnamefont {M.~R.}\ \bibnamefont {{Abernathy}}}, \bibinfo
  {author} {\bibfnamefont {F.}~\bibnamefont {{Acernese}}}, \bibinfo {author}
  {\bibfnamefont {K.}~\bibnamefont {{Ackley}}}, \bibinfo {author}
  {\bibfnamefont {C.}~\bibnamefont {{Adams}}}, \bibinfo {author} {\bibfnamefont
  {T.}~\bibnamefont {{Adams}}}, \bibinfo {author} {\bibfnamefont
  {P.}~\bibnamefont {{Addesso}}}, \bibinfo {author} {\bibfnamefont {R.~X.}\
  \bibnamefont {{Adhikari}}}, \ and\ \bibinfo {author} {\bibnamefont
  {et~al.}},\ }\href {\doibase 10.1103/PhysRevLett.116.241102} {\bibfield
  {journal} {\bibinfo  {journal} {Physical Review Letters}\ }\textbf {\bibinfo
  {volume} {116}},\ \bibinfo {eid} {241102} (\bibinfo {year}
  {2016}{\natexlab{d}})},\ \Eprint {http://arxiv.org/abs/1602.03840}
  {arXiv:1602.03840 [gr-qc]} \BibitemShut {NoStop}%
\bibitem [{\citenamefont {{Abbott}}\ \emph
  {et~al.}(2016{\natexlab{e}})\citenamefont {{Abbott}}, \citenamefont
  {{Abbott}}, \citenamefont {{Abbott}}, \citenamefont {{Abernathy}},
  \citenamefont {{Acernese}}, \citenamefont {{Ackley}}, \citenamefont
  {{Adams}}, \citenamefont {{Adams}}, \citenamefont {{Addesso}}, \citenamefont
  {{Adhikari}},\ and\ \citenamefont {et~al.}}]{gw150914astro}%
  \BibitemOpen
  \bibfield  {author} {\bibinfo {author} {\bibfnamefont {B.~P.}\ \bibnamefont
  {{Abbott}}}, \bibinfo {author} {\bibfnamefont {R.}~\bibnamefont {{Abbott}}},
  \bibinfo {author} {\bibfnamefont {T.~D.}\ \bibnamefont {{Abbott}}}, \bibinfo
  {author} {\bibfnamefont {M.~R.}\ \bibnamefont {{Abernathy}}}, \bibinfo
  {author} {\bibfnamefont {F.}~\bibnamefont {{Acernese}}}, \bibinfo {author}
  {\bibfnamefont {K.}~\bibnamefont {{Ackley}}}, \bibinfo {author}
  {\bibfnamefont {C.}~\bibnamefont {{Adams}}}, \bibinfo {author} {\bibfnamefont
  {T.}~\bibnamefont {{Adams}}}, \bibinfo {author} {\bibfnamefont
  {P.}~\bibnamefont {{Addesso}}}, \bibinfo {author} {\bibfnamefont {R.~X.}\
  \bibnamefont {{Adhikari}}}, \ and\ \bibinfo {author} {\bibnamefont
  {et~al.}},\ }\href {\doibase 10.3847/2041-8205/818/2/L22} {\bibfield
  {journal} {\bibinfo  {journal} {Astrophys. J. Supp.}\ }\textbf {\bibinfo
  {volume} {818}},\ \bibinfo {eid} {L22} (\bibinfo {year}
  {2016}{\natexlab{e}})},\ \Eprint {http://arxiv.org/abs/1602.03846}
  {arXiv:1602.03846 [astro-ph.HE]} \BibitemShut {NoStop}%
\bibitem [{\citenamefont {{Belczynski}}\ \emph {et~al.}(2010)\citenamefont
  {{Belczynski}}, \citenamefont {{Dominik}}, \citenamefont {{Bulik}},
  \citenamefont {{O'Shaughnessy}}, \citenamefont {{Fryer}},\ and\ \citenamefont
  {{Holz}}}]{2010ApJ...715L.138B}%
  \BibitemOpen
  \bibfield  {author} {\bibinfo {author} {\bibfnamefont {K.}~\bibnamefont
  {{Belczynski}}}, \bibinfo {author} {\bibfnamefont {M.}~\bibnamefont
  {{Dominik}}}, \bibinfo {author} {\bibfnamefont {T.}~\bibnamefont {{Bulik}}},
  \bibinfo {author} {\bibfnamefont {R.}~\bibnamefont {{O'Shaughnessy}}},
  \bibinfo {author} {\bibfnamefont {C.}~\bibnamefont {{Fryer}}}, \ and\
  \bibinfo {author} {\bibfnamefont {D.~E.}\ \bibnamefont {{Holz}}},\ }\href
  {\doibase 10.1088/2041-8205/715/2/L138} {\bibfield  {journal} {\bibinfo
  {journal} {Astrophys. J. Supp.}\ }\textbf {\bibinfo {volume} {715}},\
  \bibinfo {pages} {L138} (\bibinfo {year} {2010})}\BibitemShut {NoStop}%
\bibitem [{\citenamefont {{Rodriguez}}\ \emph {et~al.}(2015)\citenamefont
  {{Rodriguez}}, \citenamefont {{Morscher}}, \citenamefont {{Pattabiraman}},
  \citenamefont {{Chatterjee}}, \citenamefont {{Haster}},\ and\ \citenamefont
  {{Rasio}}}]{2015PhRvL.115e1101R}%
  \BibitemOpen
  \bibfield  {author} {\bibinfo {author} {\bibfnamefont {C.~L.}\ \bibnamefont
  {{Rodriguez}}}, \bibinfo {author} {\bibfnamefont {M.}~\bibnamefont
  {{Morscher}}}, \bibinfo {author} {\bibfnamefont {B.}~\bibnamefont
  {{Pattabiraman}}}, \bibinfo {author} {\bibfnamefont {S.}~\bibnamefont
  {{Chatterjee}}}, \bibinfo {author} {\bibfnamefont {C.-J.}\ \bibnamefont
  {{Haster}}}, \ and\ \bibinfo {author} {\bibfnamefont {F.~A.}\ \bibnamefont
  {{Rasio}}},\ }\href {\doibase 10.1103/PhysRevLett.115.051101} {\bibfield
  {journal} {\bibinfo  {journal} {Physical Review Letters}\ }\textbf {\bibinfo
  {volume} {115}},\ \bibinfo {eid} {051101} (\bibinfo {year}
  {2015})}\BibitemShut {NoStop}%
\bibitem [{\citenamefont {{The LIGO Scientific Collaboration}}\ \emph
  {et~al.}(2016{\natexlab{a}})\citenamefont {{The LIGO Scientific
  Collaboration}}, \citenamefont {{the Virgo Collaboration}}, \citenamefont
  {{Abbott}}, \citenamefont {{Abbott}}, \citenamefont {{Abbott}}, \citenamefont
  {{Abernathy}}, \citenamefont {{Acernese}}, \citenamefont {{Ackley}},
  \citenamefont {{Adams}}, \citenamefont {{Adams}},\ and\ \citenamefont
  {et~al.}}]{2016arXiv160604856T}%
  \BibitemOpen
  \bibfield  {author} {\bibinfo {author} {\bibnamefont {{The LIGO Scientific
  Collaboration}}}, \bibinfo {author} {\bibnamefont {{the Virgo
  Collaboration}}}, \bibinfo {author} {\bibfnamefont {B.~P.}\ \bibnamefont
  {{Abbott}}}, \bibinfo {author} {\bibfnamefont {R.}~\bibnamefont {{Abbott}}},
  \bibinfo {author} {\bibfnamefont {T.~D.}\ \bibnamefont {{Abbott}}}, \bibinfo
  {author} {\bibfnamefont {M.~R.}\ \bibnamefont {{Abernathy}}}, \bibinfo
  {author} {\bibfnamefont {F.}~\bibnamefont {{Acernese}}}, \bibinfo {author}
  {\bibfnamefont {K.}~\bibnamefont {{Ackley}}}, \bibinfo {author}
  {\bibfnamefont {C.}~\bibnamefont {{Adams}}}, \bibinfo {author} {\bibfnamefont
  {T.}~\bibnamefont {{Adams}}}, \ and\ \bibinfo {author} {\bibnamefont
  {et~al.}},\ }\href@noop {} {\bibfield  {journal} {\bibinfo  {journal} {ArXiv
  e-prints}\ } (\bibinfo {year} {2016}{\natexlab{a}})},\ \Eprint
  {http://arxiv.org/abs/1606.04856} {arXiv:1606.04856 [gr-qc]} \BibitemShut
  {NoStop}%
\bibitem [{\citenamefont {{The LIGO Scientific Collaboration}}\ \emph
  {et~al.}(2016{\natexlab{b}})\citenamefont {{The LIGO Scientific
  Collaboration}}, \citenamefont {{the Virgo Collaboration}}, \citenamefont
  {{Abbott}}, \citenamefont {{Abbott}}, \citenamefont {{Abbott}}, \citenamefont
  {{Abernathy}}, \citenamefont {{Acernese}}, \citenamefont {{Ackley}},
  \citenamefont {{Adams}}, \citenamefont {{Adams}},\ and\ \citenamefont
  {et~al.}}]{gw150914rate}%
  \BibitemOpen
  \bibfield  {author} {\bibinfo {author} {\bibnamefont {{The LIGO Scientific
  Collaboration}}}, \bibinfo {author} {\bibnamefont {{the Virgo
  Collaboration}}}, \bibinfo {author} {\bibfnamefont {B.~P.}\ \bibnamefont
  {{Abbott}}}, \bibinfo {author} {\bibfnamefont {R.}~\bibnamefont {{Abbott}}},
  \bibinfo {author} {\bibfnamefont {T.~D.}\ \bibnamefont {{Abbott}}}, \bibinfo
  {author} {\bibfnamefont {M.~R.}\ \bibnamefont {{Abernathy}}}, \bibinfo
  {author} {\bibfnamefont {F.}~\bibnamefont {{Acernese}}}, \bibinfo {author}
  {\bibfnamefont {K.}~\bibnamefont {{Ackley}}}, \bibinfo {author}
  {\bibfnamefont {C.}~\bibnamefont {{Adams}}}, \bibinfo {author} {\bibfnamefont
  {T.}~\bibnamefont {{Adams}}}, \ and\ \bibinfo {author} {\bibnamefont
  {et~al.}},\ }\href@noop {} {\bibfield  {journal} {\bibinfo  {journal} {ArXiv
  e-prints}\ } (\bibinfo {year} {2016}{\natexlab{b}})},\ \Eprint
  {http://arxiv.org/abs/1602.03842} {arXiv:1602.03842 [astro-ph.HE]}
  \BibitemShut {NoStop}%
\bibitem [{\citenamefont {{Abbott}}\ \emph
  {et~al.}(2016{\natexlab{f}})\citenamefont {{Abbott}}, \citenamefont
  {{Abbott}}, \citenamefont {{Abbott}}, \citenamefont {{Abernathy}},
  \citenamefont {{Acernese}}, \citenamefont {{Ackley}}, \citenamefont
  {{Adams}}, \citenamefont {{Adams}}, \citenamefont {{Addesso}}, \citenamefont
  {{Adhikari}},\ and\ \citenamefont {et~al.}}]{gw150914sgwb}%
  \BibitemOpen
  \bibfield  {author} {\bibinfo {author} {\bibfnamefont {B.~P.}\ \bibnamefont
  {{Abbott}}}, \bibinfo {author} {\bibfnamefont {R.}~\bibnamefont {{Abbott}}},
  \bibinfo {author} {\bibfnamefont {T.~D.}\ \bibnamefont {{Abbott}}}, \bibinfo
  {author} {\bibfnamefont {M.~R.}\ \bibnamefont {{Abernathy}}}, \bibinfo
  {author} {\bibfnamefont {F.}~\bibnamefont {{Acernese}}}, \bibinfo {author}
  {\bibfnamefont {K.}~\bibnamefont {{Ackley}}}, \bibinfo {author}
  {\bibfnamefont {C.}~\bibnamefont {{Adams}}}, \bibinfo {author} {\bibfnamefont
  {T.}~\bibnamefont {{Adams}}}, \bibinfo {author} {\bibfnamefont
  {P.}~\bibnamefont {{Addesso}}}, \bibinfo {author} {\bibfnamefont {R.~X.}\
  \bibnamefont {{Adhikari}}}, \ and\ \bibinfo {author} {\bibnamefont
  {et~al.}},\ }\href {\doibase 10.1103/PhysRevLett.116.131102} {\bibfield
  {journal} {\bibinfo  {journal} {Physical Review Letters}\ }\textbf {\bibinfo
  {volume} {116}},\ \bibinfo {eid} {131102} (\bibinfo {year}
  {2016}{\natexlab{f}})}\BibitemShut {NoStop}%
\bibitem [{\citenamefont {{Zhu}}\ \emph {et~al.}(2011)\citenamefont {{Zhu}},
  \citenamefont {{Howell}}, \citenamefont {{Regimbau}}, \citenamefont
  {{Blair}},\ and\ \citenamefont {{Zhu}}}]{2011ApJ...739...86Z}%
  \BibitemOpen
  \bibfield  {author} {\bibinfo {author} {\bibfnamefont {X.-J.}\ \bibnamefont
  {{Zhu}}}, \bibinfo {author} {\bibfnamefont {E.}~\bibnamefont {{Howell}}},
  \bibinfo {author} {\bibfnamefont {T.}~\bibnamefont {{Regimbau}}}, \bibinfo
  {author} {\bibfnamefont {D.}~\bibnamefont {{Blair}}}, \ and\ \bibinfo
  {author} {\bibfnamefont {Z.-H.}\ \bibnamefont {{Zhu}}},\ }\href {\doibase
  10.1088/0004-637X/739/2/86} {\bibfield  {journal} {\bibinfo  {journal}
  {\apj}\ }\textbf {\bibinfo {volume} {739}},\ \bibinfo {eid} {86} (\bibinfo
  {year} {2011})}\BibitemShut {NoStop}%
\bibitem [{\citenamefont {{Rosado}}(2011)}]{2011PhRvD..84h4004R}%
  \BibitemOpen
  \bibfield  {author} {\bibinfo {author} {\bibfnamefont {P.~A.}\ \bibnamefont
  {{Rosado}}},\ }\href {\doibase 10.1103/PhysRevD.84.084004} {\bibfield
  {journal} {\bibinfo  {journal} {\prd}\ }\textbf {\bibinfo {volume} {84}},\
  \bibinfo {eid} {084004} (\bibinfo {year} {2011})}\BibitemShut {NoStop}%
\bibitem [{\citenamefont {{Marassi}}\ \emph {et~al.}(2011)\citenamefont
  {{Marassi}}, \citenamefont {{Schneider}}, \citenamefont {{Corvino}},
  \citenamefont {{Ferrari}},\ and\ \citenamefont
  {{Zwart}}}]{2011PhRvD..84l4037M}%
  \BibitemOpen
  \bibfield  {author} {\bibinfo {author} {\bibfnamefont {S.}~\bibnamefont
  {{Marassi}}}, \bibinfo {author} {\bibfnamefont {R.}~\bibnamefont
  {{Schneider}}}, \bibinfo {author} {\bibfnamefont {G.}~\bibnamefont
  {{Corvino}}}, \bibinfo {author} {\bibfnamefont {V.}~\bibnamefont
  {{Ferrari}}}, \ and\ \bibinfo {author} {\bibfnamefont {S.~P.}\ \bibnamefont
  {{Zwart}}},\ }\href {\doibase 10.1103/PhysRevD.84.124037} {\bibfield
  {journal} {\bibinfo  {journal} {\prd}\ }\textbf {\bibinfo {volume} {84}},\
  \bibinfo {eid} {124037} (\bibinfo {year} {2011})}\BibitemShut {NoStop}%
\bibitem [{\citenamefont {{Wu}}\ \emph {et~al.}(2012)\citenamefont {{Wu}},
  \citenamefont {{Mandic}},\ and\ \citenamefont
  {{Regimbau}}}]{2012PhRvD..85j4024W}%
  \BibitemOpen
  \bibfield  {author} {\bibinfo {author} {\bibfnamefont {C.}~\bibnamefont
  {{Wu}}}, \bibinfo {author} {\bibfnamefont {V.}~\bibnamefont {{Mandic}}}, \
  and\ \bibinfo {author} {\bibfnamefont {T.}~\bibnamefont {{Regimbau}}},\
  }\href {\doibase 10.1103/PhysRevD.85.104024} {\bibfield  {journal} {\bibinfo
  {journal} {\prd}\ }\textbf {\bibinfo {volume} {85}},\ \bibinfo {eid} {104024}
  (\bibinfo {year} {2012})}\BibitemShut {NoStop}%
\bibitem [{\citenamefont {{Zhu}}\ \emph {et~al.}(2013)\citenamefont {{Zhu}},
  \citenamefont {{Howell}}, \citenamefont {{Blair}},\ and\ \citenamefont
  {{Zhu}}}]{2013MNRAS.431..882Z}%
  \BibitemOpen
  \bibfield  {author} {\bibinfo {author} {\bibfnamefont {X.-J.}\ \bibnamefont
  {{Zhu}}}, \bibinfo {author} {\bibfnamefont {E.~J.}\ \bibnamefont {{Howell}}},
  \bibinfo {author} {\bibfnamefont {D.~G.}\ \bibnamefont {{Blair}}}, \ and\
  \bibinfo {author} {\bibfnamefont {Z.-H.}\ \bibnamefont {{Zhu}}},\ }\href
  {\doibase 10.1093/mnras/stt207} {\bibfield  {journal} {\bibinfo  {journal}
  {Monthly Notices of the Royal astronomical Society}\ }\textbf {\bibinfo
  {volume} {431}},\ \bibinfo {pages} {882} (\bibinfo {year}
  {2013})}\BibitemShut {NoStop}%
\bibitem [{\citenamefont {{Kowalska-Leszczynska}}\ \emph
  {et~al.}(2015)\citenamefont {{Kowalska-Leszczynska}}, \citenamefont
  {{Regimbau}}, \citenamefont {{Bulik}}, \citenamefont {{Dominik}},\ and\
  \citenamefont {{Belczynski}}}]{2015A&A...574A..58K}%
  \BibitemOpen
  \bibfield  {author} {\bibinfo {author} {\bibfnamefont {I.}~\bibnamefont
  {{Kowalska-Leszczynska}}}, \bibinfo {author} {\bibfnamefont {T.}~\bibnamefont
  {{Regimbau}}}, \bibinfo {author} {\bibfnamefont {T.}~\bibnamefont {{Bulik}}},
  \bibinfo {author} {\bibfnamefont {M.}~\bibnamefont {{Dominik}}}, \ and\
  \bibinfo {author} {\bibfnamefont {K.}~\bibnamefont {{Belczynski}}},\ }\href
  {\doibase 10.1051/0004-6361/201424417} {\bibfield  {journal} {\bibinfo
  {journal} {Astronomy and Astrophysics}\ }\textbf {\bibinfo {volume} {574}},\
  \bibinfo {eid} {A58} (\bibinfo {year} {2015})}\BibitemShut {NoStop}%
\bibitem [{\citenamefont {{Meacher}}\ \emph {et~al.}(2015)\citenamefont
  {{Meacher}}, \citenamefont {{Coughlin}}, \citenamefont {{Morris}},
  \citenamefont {{Regimbau}}, \citenamefont {{Christensen}}, \citenamefont
  {{Kandhasamy}}, \citenamefont {{Mandic}}, \citenamefont {{Romano}},\ and\
  \citenamefont {{Thrane}}}]{2015PhRvD..92f3002M}%
  \BibitemOpen
  \bibfield  {author} {\bibinfo {author} {\bibfnamefont {D.}~\bibnamefont
  {{Meacher}}}, \bibinfo {author} {\bibfnamefont {M.}~\bibnamefont
  {{Coughlin}}}, \bibinfo {author} {\bibfnamefont {S.}~\bibnamefont
  {{Morris}}}, \bibinfo {author} {\bibfnamefont {T.}~\bibnamefont
  {{Regimbau}}}, \bibinfo {author} {\bibfnamefont {N.}~\bibnamefont
  {{Christensen}}}, \bibinfo {author} {\bibfnamefont {S.}~\bibnamefont
  {{Kandhasamy}}}, \bibinfo {author} {\bibfnamefont {V.}~\bibnamefont
  {{Mandic}}}, \bibinfo {author} {\bibfnamefont {J.~D.}\ \bibnamefont
  {{Romano}}}, \ and\ \bibinfo {author} {\bibfnamefont {E.}~\bibnamefont
  {{Thrane}}},\ }\href {\doibase 10.1103/PhysRevD.92.063002} {\bibfield
  {journal} {\bibinfo  {journal} {Phys. Rev. D}\ }\textbf {\bibinfo {volume}
  {92}},\ \bibinfo {eid} {063002} (\bibinfo {year} {2015})},\ \Eprint
  {http://arxiv.org/abs/1506.06744} {arXiv:1506.06744 [astro-ph.HE]}
  \BibitemShut {NoStop}%
\bibitem [{\citenamefont {{Dvorkin}}\ \emph {et~al.}(2016)\citenamefont
  {{Dvorkin}}, \citenamefont {{Vangioni}}, \citenamefont {{Silk}},
  \citenamefont {{Uzan}},\ and\ \citenamefont {{Olive}}}]{2016arXiv160404288D}%
  \BibitemOpen
  \bibfield  {author} {\bibinfo {author} {\bibfnamefont {I.}~\bibnamefont
  {{Dvorkin}}}, \bibinfo {author} {\bibfnamefont {E.}~\bibnamefont
  {{Vangioni}}}, \bibinfo {author} {\bibfnamefont {J.}~\bibnamefont {{Silk}}},
  \bibinfo {author} {\bibfnamefont {J.-P.}\ \bibnamefont {{Uzan}}}, \ and\
  \bibinfo {author} {\bibfnamefont {K.~A.}\ \bibnamefont {{Olive}}},\
  }\href@noop {} {\bibfield  {journal} {\bibinfo  {journal} {ArXiv e-prints}\ }
  (\bibinfo {year} {2016})},\ \Eprint {http://arxiv.org/abs/1604.04288}
  {arXiv:1604.04288 [astro-ph.HE]} \BibitemShut {NoStop}%
\bibitem [{\citenamefont {{Nakazato}}\ \emph {et~al.}(2016)\citenamefont
  {{Nakazato}}, \citenamefont {{Niino}},\ and\ \citenamefont
  {{Sago}}}]{2016arXiv160502146N}%
  \BibitemOpen
  \bibfield  {author} {\bibinfo {author} {\bibfnamefont {K.}~\bibnamefont
  {{Nakazato}}}, \bibinfo {author} {\bibfnamefont {Y.}~\bibnamefont {{Niino}}},
  \ and\ \bibinfo {author} {\bibfnamefont {N.}~\bibnamefont {{Sago}}},\
  }\href@noop {} {\bibfield  {journal} {\bibinfo  {journal} {ArXiv e-prints}\ }
  (\bibinfo {year} {2016})},\ \Eprint {http://arxiv.org/abs/1605.02146}
  {arXiv:1605.02146 [astro-ph.HE]} \BibitemShut {NoStop}%
\bibitem [{\citenamefont {{Regimbau}}\ \emph
  {et~al.}(2012{\natexlab{b}})\citenamefont {{Regimbau}}, \citenamefont
  {{Dent}}, \citenamefont {{Del Pozzo}}, \citenamefont {{Giampanis}},
  \citenamefont {{Li}}, \citenamefont {{Robinson}}, \citenamefont {{Van Den
  Broeck}}, \citenamefont {{Meacher}}, \citenamefont {{Rodriguez}},
  \citenamefont {{Sathyaprakash}},\ and\ \citenamefont
  {{W{\'o}jcik}}}]{ETMDC1}%
  \BibitemOpen
  \bibfield  {author} {\bibinfo {author} {\bibfnamefont {T.}~\bibnamefont
  {{Regimbau}}}, \bibinfo {author} {\bibfnamefont {T.}~\bibnamefont {{Dent}}},
  \bibinfo {author} {\bibfnamefont {W.}~\bibnamefont {{Del Pozzo}}}, \bibinfo
  {author} {\bibfnamefont {S.}~\bibnamefont {{Giampanis}}}, \bibinfo {author}
  {\bibfnamefont {T.~G.~F.}\ \bibnamefont {{Li}}}, \bibinfo {author}
  {\bibfnamefont {C.}~\bibnamefont {{Robinson}}}, \bibinfo {author}
  {\bibfnamefont {C.}~\bibnamefont {{Van Den Broeck}}}, \bibinfo {author}
  {\bibfnamefont {D.}~\bibnamefont {{Meacher}}}, \bibinfo {author}
  {\bibfnamefont {C.}~\bibnamefont {{Rodriguez}}}, \bibinfo {author}
  {\bibfnamefont {B.~S.}\ \bibnamefont {{Sathyaprakash}}}, \ and\ \bibinfo
  {author} {\bibfnamefont {K.}~\bibnamefont {{W{\'o}jcik}}},\ }\href {\doibase
  10.1103/PhysRevD.86.122001} {\bibfield  {journal} {\bibinfo  {journal}
  {\prd}\ }\textbf {\bibinfo {volume} {86}},\ \bibinfo {eid} {122001} (\bibinfo
  {year} {2012}{\natexlab{b}})},\ \Eprint {http://arxiv.org/abs/1201.3563}
  {arXiv:1201.3563 [gr-qc]} \BibitemShut {NoStop}%
\bibitem [{\citenamefont {{Regimbau}}\ \emph {et~al.}(2014)\citenamefont
  {{Regimbau}}, \citenamefont {{Meacher}},\ and\ \citenamefont
  {{Coughlin}}}]{ETMDC2}%
  \BibitemOpen
  \bibfield  {author} {\bibinfo {author} {\bibfnamefont {T.}~\bibnamefont
  {{Regimbau}}}, \bibinfo {author} {\bibfnamefont {D.}~\bibnamefont
  {{Meacher}}}, \ and\ \bibinfo {author} {\bibfnamefont {M.}~\bibnamefont
  {{Coughlin}}},\ }\href {\doibase 10.1103/PhysRevD.89.084046} {\bibfield
  {journal} {\bibinfo  {journal} {\prd}\ }\textbf {\bibinfo {volume} {89}},\
  \bibinfo {eid} {084046} (\bibinfo {year} {2014})},\ \Eprint
  {http://arxiv.org/abs/1404.1134} {arXiv:1404.1134} \BibitemShut {NoStop}%
\bibitem [{\citenamefont {{Vangioni}}\ \emph {et~al.}(2015)\citenamefont
  {{Vangioni}}, \citenamefont {{Olive}}, \citenamefont {{Prestegard}},
  \citenamefont {{Silk}}, \citenamefont {{Petitjean}},\ and\ \citenamefont
  {{Mandic}}}]{2015MNRAS.447.2575V}%
  \BibitemOpen
  \bibfield  {author} {\bibinfo {author} {\bibfnamefont {E.}~\bibnamefont
  {{Vangioni}}}, \bibinfo {author} {\bibfnamefont {K.~A.}\ \bibnamefont
  {{Olive}}}, \bibinfo {author} {\bibfnamefont {T.}~\bibnamefont
  {{Prestegard}}}, \bibinfo {author} {\bibfnamefont {J.}~\bibnamefont
  {{Silk}}}, \bibinfo {author} {\bibfnamefont {P.}~\bibnamefont {{Petitjean}}},
  \ and\ \bibinfo {author} {\bibfnamefont {V.}~\bibnamefont {{Mandic}}},\
  }\href {\doibase 10.1093/mnras/stu2600} {\bibfield  {journal} {\bibinfo
  {journal} {Mon. Not. Roy. Astr. Soc.}\ }\textbf {\bibinfo {volume} {447}},\
  \bibinfo {pages} {2575} (\bibinfo {year} {2015})}\BibitemShut {NoStop}%
\bibitem [{\citenamefont {Kistler}\ \emph {et~al.}(2013)\citenamefont
  {Kistler}, \citenamefont {Yuksel},\ and\ \citenamefont {Hopkins}}]{kistler}%
  \BibitemOpen
  \bibfield  {author} {\bibinfo {author} {\bibfnamefont {M.}~\bibnamefont
  {Kistler}}, \bibinfo {author} {\bibfnamefont {H.}~\bibnamefont {Yuksel}}, \
  and\ \bibinfo {author} {\bibfnamefont {A.}~\bibnamefont {Hopkins}},\
  }\href@noop {} {\  (\bibinfo {year} {2013})},\ \Eprint
  {http://arxiv.org/abs/1305.1630} {arXiv:1305.1630} \BibitemShut {NoStop}%
\bibitem [{\citenamefont {Trenti}\ \emph {et~al.}(2013)\citenamefont {Trenti},
  \citenamefont {Perna},\ and\ \citenamefont {Tacchella}}]{trenti}%
  \BibitemOpen
  \bibfield  {author} {\bibinfo {author} {\bibfnamefont {M.}~\bibnamefont
  {Trenti}}, \bibinfo {author} {\bibfnamefont {R.}~\bibnamefont {Perna}}, \
  and\ \bibinfo {author} {\bibfnamefont {S.}~\bibnamefont {Tacchella}},\
  }\href@noop {} {\bibfield  {journal} {\bibinfo  {journal} {Astrophys. J.
  Supp.}\ }\textbf {\bibinfo {volume} {773}},\ \bibinfo {pages} {22} (\bibinfo
  {year} {2013})}\BibitemShut {NoStop}%
\bibitem [{\citenamefont {Behroozi}\ and\ \citenamefont
  {Silk}(2015)}]{behroozi}%
  \BibitemOpen
  \bibfield  {author} {\bibinfo {author} {\bibfnamefont {P.}~\bibnamefont
  {Behroozi}}\ and\ \bibinfo {author} {\bibfnamefont {J.}~\bibnamefont
  {Silk}},\ }\href@noop {} {\bibfield  {journal} {\bibinfo  {journal}
  {Astrophys. J.}\ }\textbf {\bibinfo {volume} {799}},\ \bibinfo {pages} {32}
  (\bibinfo {year} {2015})}\BibitemShut {NoStop}%
\bibitem [{\citenamefont {{Dvorkin}}\ \emph {et~al.}(2015)\citenamefont
  {{Dvorkin}}, \citenamefont {{Silk}}, \citenamefont {{Vangioni}},
  \citenamefont {{Petitjean}},\ and\ \citenamefont
  {{Olive}}}]{2015MNRAS.452L..36D}%
  \BibitemOpen
  \bibfield  {author} {\bibinfo {author} {\bibfnamefont {I.}~\bibnamefont
  {{Dvorkin}}}, \bibinfo {author} {\bibfnamefont {J.}~\bibnamefont {{Silk}}},
  \bibinfo {author} {\bibfnamefont {E.}~\bibnamefont {{Vangioni}}}, \bibinfo
  {author} {\bibfnamefont {P.}~\bibnamefont {{Petitjean}}}, \ and\ \bibinfo
  {author} {\bibfnamefont {K.~A.}\ \bibnamefont {{Olive}}},\ }\href {\doibase
  10.1093/mnrasl/slv085} {\bibfield  {journal} {\bibinfo  {journal} {Mon. Not.
  Roy. Astr. Soc.}\ }\textbf {\bibinfo {volume} {452}},\ \bibinfo {pages} {L36}
  (\bibinfo {year} {2015})},\ \Eprint {http://arxiv.org/abs/1506.06761}
  {arXiv:1506.06761} \BibitemShut {NoStop}%
\bibitem [{\citenamefont {{Madau}}\ and\ \citenamefont
  {{Dickinson}}(2014)}]{2014ARAA..52..415M}%
  \BibitemOpen
  \bibfield  {author} {\bibinfo {author} {\bibfnamefont {P.}~\bibnamefont
  {{Madau}}}\ and\ \bibinfo {author} {\bibfnamefont {M.}~\bibnamefont
  {{Dickinson}}},\ }\href@noop {} {\bibfield  {journal} {\bibinfo  {journal}
  {AARA}\ }\textbf {\bibinfo {volume} {52}},\ \bibinfo {pages} {415} (\bibinfo
  {year} {2014})}\BibitemShut {NoStop}%
\bibitem [{\citenamefont {{Belczynski}}\ \emph {et~al.}(2016)\citenamefont
  {{Belczynski}}, \citenamefont {{Repetto}}, \citenamefont {{Holz}},
  \citenamefont {{O'Shaughnessy}}, \citenamefont {{Bulik}}, \citenamefont
  {{Berti}}, \citenamefont {{Fryer}},\ and\ \citenamefont
  {{Dominik}}}]{gw150914chris}%
  \BibitemOpen
  \bibfield  {author} {\bibinfo {author} {\bibfnamefont {K.}~\bibnamefont
  {{Belczynski}}}, \bibinfo {author} {\bibfnamefont {S.}~\bibnamefont
  {{Repetto}}}, \bibinfo {author} {\bibfnamefont {D.~E.}\ \bibnamefont
  {{Holz}}}, \bibinfo {author} {\bibfnamefont {R.}~\bibnamefont
  {{O'Shaughnessy}}}, \bibinfo {author} {\bibfnamefont {T.}~\bibnamefont
  {{Bulik}}}, \bibinfo {author} {\bibfnamefont {E.}~\bibnamefont {{Berti}}},
  \bibinfo {author} {\bibfnamefont {C.}~\bibnamefont {{Fryer}}}, \ and\
  \bibinfo {author} {\bibfnamefont {M.}~\bibnamefont {{Dominik}}},\ }\href
  {\doibase 10.3847/0004-637X/819/2/108} {\bibfield  {journal} {\bibinfo
  {journal} {\apj}\ }\textbf {\bibinfo {volume} {819}},\ \bibinfo {eid} {108}
  (\bibinfo {year} {2016})},\ \Eprint {http://arxiv.org/abs/1510.04615}
  {arXiv:1510.04615 [astro-ph.HE]} \BibitemShut {NoStop}%
\bibitem [{\citenamefont {{Belczynski}}\ \emph {et~al.}(2002)\citenamefont
  {{Belczynski}}, \citenamefont {{Kalogera}},\ and\ \citenamefont
  {{Bulik}}}]{2002ApJ...572..407B}%
  \BibitemOpen
  \bibfield  {author} {\bibinfo {author} {\bibfnamefont {K.}~\bibnamefont
  {{Belczynski}}}, \bibinfo {author} {\bibfnamefont {V.}~\bibnamefont
  {{Kalogera}}}, \ and\ \bibinfo {author} {\bibfnamefont {T.}~\bibnamefont
  {{Bulik}}},\ }\href {\doibase 10.1086/340304} {\bibfield  {journal} {\bibinfo
   {journal} {\apj}\ }\textbf {\bibinfo {volume} {572}},\ \bibinfo {pages}
  {407} (\bibinfo {year} {2002})}\BibitemShut {NoStop}%
\bibitem [{\citenamefont {{Ando}}(2004)}]{2004JCAP...06..007A}%
  \BibitemOpen
  \bibfield  {author} {\bibinfo {author} {\bibfnamefont {S.}~\bibnamefont
  {{Ando}}},\ }\href {\doibase 10.1088/1475-7516/2004/06/007} {\bibfield
  {journal} {\bibinfo  {journal} {Journal of Cosmology and Astroparticle
  Physic}\ }\textbf {\bibinfo {volume} {6}},\ \bibinfo {eid} {007} (\bibinfo
  {year} {2004})}\BibitemShut {NoStop}%
\bibitem [{\citenamefont {{Belczynski}}\ \emph {et~al.}(2006)\citenamefont
  {{Belczynski}}, \citenamefont {{Perna}}, \citenamefont {{Bulik}},
  \citenamefont {{Kalogera}}, \citenamefont {{Ivanova}},\ and\ \citenamefont
  {{Lamb}}}]{2006ApJ...648.1110B}%
  \BibitemOpen
  \bibfield  {author} {\bibinfo {author} {\bibfnamefont {K.}~\bibnamefont
  {{Belczynski}}}, \bibinfo {author} {\bibfnamefont {R.}~\bibnamefont
  {{Perna}}}, \bibinfo {author} {\bibfnamefont {T.}~\bibnamefont {{Bulik}}},
  \bibinfo {author} {\bibfnamefont {V.}~\bibnamefont {{Kalogera}}}, \bibinfo
  {author} {\bibfnamefont {N.}~\bibnamefont {{Ivanova}}}, \ and\ \bibinfo
  {author} {\bibfnamefont {D.~Q.}\ \bibnamefont {{Lamb}}},\ }\href {\doibase
  10.1086/505169} {\bibfield  {journal} {\bibinfo  {journal} {\apj}\ }\textbf
  {\bibinfo {volume} {648}},\ \bibinfo {pages} {1110} (\bibinfo {year}
  {2006})}\BibitemShut {NoStop}%
\bibitem [{\citenamefont {{de Freitas Pacheco}}\ \emph
  {et~al.}(2006)\citenamefont {{de Freitas Pacheco}}, \citenamefont
  {{Regimbau}}, \citenamefont {{Vincent}},\ and\ \citenamefont
  {{Spallicci}}}]{2006IJMPD..15..235D}%
  \BibitemOpen
  \bibfield  {author} {\bibinfo {author} {\bibfnamefont {J.~A.}\ \bibnamefont
  {{de Freitas Pacheco}}}, \bibinfo {author} {\bibfnamefont {T.}~\bibnamefont
  {{Regimbau}}}, \bibinfo {author} {\bibfnamefont {S.}~\bibnamefont
  {{Vincent}}}, \ and\ \bibinfo {author} {\bibfnamefont {A.}~\bibnamefont
  {{Spallicci}}},\ }\href {\doibase 10.1142/S0218271806007699} {\bibfield
  {journal} {\bibinfo  {journal} {International Journal of Modern Physics D}\
  }\textbf {\bibinfo {volume} {15}},\ \bibinfo {pages} {235} (\bibinfo {year}
  {2006})}\BibitemShut {NoStop}%
\bibitem [{\citenamefont {{Berger}}\ \emph {et~al.}(2007)\citenamefont
  {{Berger}}, \citenamefont {{Fox}}, \citenamefont {{Price}}, \citenamefont
  {{Nakar}}, \citenamefont {{Gal-Yam}}, \citenamefont {{Holz}}, \citenamefont
  {{Schmidt}}, \citenamefont {{Cucchiara}}, \citenamefont {{Cenko}},
  \citenamefont {{Kulkarni}}, \citenamefont {{Soderberg}}, \citenamefont
  {{Frail}}, \citenamefont {{Penprase}}, \citenamefont {{Rau}}, \citenamefont
  {{Ofek}}, \citenamefont {{Burnell}}, \citenamefont {{Cameron}}, \citenamefont
  {{Cowie}}, \citenamefont {{Dopita}}, \citenamefont {{Hook}}, \citenamefont
  {{Peterson}}, \citenamefont {{Podsiadlowski}}, \citenamefont {{Roth}},
  \citenamefont {{Rutledge}}, \citenamefont {{Sheppard}},\ and\ \citenamefont
  {{Songaila}}}]{2007ApJ...664.1000B}%
  \BibitemOpen
  \bibfield  {author} {\bibinfo {author} {\bibfnamefont {E.}~\bibnamefont
  {{Berger}}}, \bibinfo {author} {\bibfnamefont {D.~B.}\ \bibnamefont {{Fox}}},
  \bibinfo {author} {\bibfnamefont {P.~A.}\ \bibnamefont {{Price}}}, \bibinfo
  {author} {\bibfnamefont {E.}~\bibnamefont {{Nakar}}}, \bibinfo {author}
  {\bibfnamefont {A.}~\bibnamefont {{Gal-Yam}}}, \bibinfo {author}
  {\bibfnamefont {D.~E.}\ \bibnamefont {{Holz}}}, \bibinfo {author}
  {\bibfnamefont {B.~P.}\ \bibnamefont {{Schmidt}}}, \bibinfo {author}
  {\bibfnamefont {A.}~\bibnamefont {{Cucchiara}}}, \bibinfo {author}
  {\bibfnamefont {S.~B.}\ \bibnamefont {{Cenko}}}, \bibinfo {author}
  {\bibfnamefont {S.~R.}\ \bibnamefont {{Kulkarni}}}, \bibinfo {author}
  {\bibfnamefont {A.~M.}\ \bibnamefont {{Soderberg}}}, \bibinfo {author}
  {\bibfnamefont {D.~A.}\ \bibnamefont {{Frail}}}, \bibinfo {author}
  {\bibfnamefont {B.~E.}\ \bibnamefont {{Penprase}}}, \bibinfo {author}
  {\bibfnamefont {A.}~\bibnamefont {{Rau}}}, \bibinfo {author} {\bibfnamefont
  {E.}~\bibnamefont {{Ofek}}}, \bibinfo {author} {\bibfnamefont {S.~J.~B.}\
  \bibnamefont {{Burnell}}}, \bibinfo {author} {\bibfnamefont {P.~B.}\
  \bibnamefont {{Cameron}}}, \bibinfo {author} {\bibfnamefont {L.~L.}\
  \bibnamefont {{Cowie}}}, \bibinfo {author} {\bibfnamefont {M.~A.}\
  \bibnamefont {{Dopita}}}, \bibinfo {author} {\bibfnamefont {I.}~\bibnamefont
  {{Hook}}}, \bibinfo {author} {\bibfnamefont {B.~A.}\ \bibnamefont
  {{Peterson}}}, \bibinfo {author} {\bibfnamefont {P.}~\bibnamefont
  {{Podsiadlowski}}}, \bibinfo {author} {\bibfnamefont {K.~C.}\ \bibnamefont
  {{Roth}}}, \bibinfo {author} {\bibfnamefont {R.~E.}\ \bibnamefont
  {{Rutledge}}}, \bibinfo {author} {\bibfnamefont {S.~S.}\ \bibnamefont
  {{Sheppard}}}, \ and\ \bibinfo {author} {\bibfnamefont {A.}~\bibnamefont
  {{Songaila}}},\ }\href {\doibase 10.1086/518762} {\bibfield  {journal}
  {\bibinfo  {journal} {\apj}\ }\textbf {\bibinfo {volume} {664}},\ \bibinfo
  {pages} {1000} (\bibinfo {year} {2007})}\BibitemShut {NoStop}%
\bibitem [{\citenamefont {{Nakar}}(2007)}]{2007PhR...442..166N}%
  \BibitemOpen
  \bibfield  {author} {\bibinfo {author} {\bibfnamefont {E.}~\bibnamefont
  {{Nakar}}},\ }\href {\doibase 10.1016/j.physrep.2007.02.005} {\bibfield
  {journal} {\bibinfo  {journal} {Phys. Rep.}\ }\textbf {\bibinfo {volume}
  {442}},\ \bibinfo {pages} {166} (\bibinfo {year} {2007})}\BibitemShut
  {NoStop}%
\bibitem [{\citenamefont {{O'Shaughnessy}}\ \emph {et~al.}(2008)\citenamefont
  {{O'Shaughnessy}}, \citenamefont {{Belczynski}},\ and\ \citenamefont
  {{Kalogera}}}]{2008ApJ...675..566O}%
  \BibitemOpen
  \bibfield  {author} {\bibinfo {author} {\bibfnamefont {R.}~\bibnamefont
  {{O'Shaughnessy}}}, \bibinfo {author} {\bibfnamefont {K.}~\bibnamefont
  {{Belczynski}}}, \ and\ \bibinfo {author} {\bibfnamefont {V.}~\bibnamefont
  {{Kalogera}}},\ }\href {\doibase 10.1086/526334} {\bibfield  {journal}
  {\bibinfo  {journal} {\apj}\ }\textbf {\bibinfo {volume} {675}},\ \bibinfo
  {pages} {566} (\bibinfo {year} {2008})}\BibitemShut {NoStop}%
\bibitem [{\citenamefont {{Dominik}}\ \emph {et~al.}(2012)\citenamefont
  {{Dominik}}, \citenamefont {{Belczynski}}, \citenamefont {{Fryer}},
  \citenamefont {{Holz}}, \citenamefont {{Berti}}, \citenamefont {{Bulik}},
  \citenamefont {{Mandel}},\ and\ \citenamefont
  {{O'Shaughnessy}}}]{2012ApJ...759...52D}%
  \BibitemOpen
  \bibfield  {author} {\bibinfo {author} {\bibfnamefont {M.}~\bibnamefont
  {{Dominik}}}, \bibinfo {author} {\bibfnamefont {K.}~\bibnamefont
  {{Belczynski}}}, \bibinfo {author} {\bibfnamefont {C.}~\bibnamefont
  {{Fryer}}}, \bibinfo {author} {\bibfnamefont {D.~E.}\ \bibnamefont {{Holz}}},
  \bibinfo {author} {\bibfnamefont {E.}~\bibnamefont {{Berti}}}, \bibinfo
  {author} {\bibfnamefont {T.}~\bibnamefont {{Bulik}}}, \bibinfo {author}
  {\bibfnamefont {I.}~\bibnamefont {{Mandel}}}, \ and\ \bibinfo {author}
  {\bibfnamefont {R.}~\bibnamefont {{O'Shaughnessy}}},\ }\href {\doibase
  10.1088/0004-637X/759/1/52} {\bibfield  {journal} {\bibinfo  {journal}
  {\apj}\ }\textbf {\bibinfo {volume} {759}},\ \bibinfo {eid} {52} (\bibinfo
  {year} {2012})}\BibitemShut {NoStop}%
\bibitem [{\citenamefont {Dominik}\ \emph {et~al.}(2013)\citenamefont
  {Dominik}, \citenamefont {Belczynski}, \citenamefont {Fryer}, \citenamefont
  {Holz}, \citenamefont {Berti}, \citenamefont {Bulik}, \citenamefont
  {Mandel},\ and\ \citenamefont {O'Shaughnessy}}]{dominik}%
  \BibitemOpen
  \bibfield  {author} {\bibinfo {author} {\bibfnamefont {M.}~\bibnamefont
  {Dominik}}, \bibinfo {author} {\bibfnamefont {K.}~\bibnamefont {Belczynski}},
  \bibinfo {author} {\bibfnamefont {C.}~\bibnamefont {Fryer}}, \bibinfo
  {author} {\bibfnamefont {D.~E.}\ \bibnamefont {Holz}}, \bibinfo {author}
  {\bibfnamefont {E.}~\bibnamefont {Berti}}, \bibinfo {author} {\bibfnamefont
  {T.}~\bibnamefont {Bulik}}, \bibinfo {author} {\bibfnamefont
  {I.}~\bibnamefont {Mandel}}, \ and\ \bibinfo {author} {\bibfnamefont
  {R.}~\bibnamefont {O'Shaughnessy}},\ }\href@noop {} {\bibfield  {journal}
  {\bibinfo  {journal} {Astrophys. J.}\ }\textbf {\bibinfo {volume} {779}},\
  \bibinfo {pages} {72} (\bibinfo {year} {2013})}\BibitemShut {NoStop}%
\bibitem [{\citenamefont {{Ajith}}(2011)}]{2011PhRvD..84h4037A}%
  \BibitemOpen
  \bibfield  {author} {\bibinfo {author} {\bibfnamefont {P.}~\bibnamefont
  {{Ajith}}},\ }\href {\doibase 10.1103/PhysRevD.84.084037} {\bibfield
  {journal} {\bibinfo  {journal} {\prd}\ }\textbf {\bibinfo {volume} {84}},\
  \bibinfo {eid} {084037} (\bibinfo {year} {2011})},\ \Eprint
  {http://arxiv.org/abs/1107.1267} {arXiv:1107.1267 [gr-qc]} \BibitemShut
  {NoStop}%
\bibitem [{\citenamefont {Aasi}\ \emph {et~al.}(2013)\citenamefont {Aasi} \emph
  {et~al.}}]{ObsScenario}%
  \BibitemOpen
  \bibfield  {author} {\bibinfo {author} {\bibfnamefont {J.}~\bibnamefont
  {Aasi}} \emph {et~al.} (\bibinfo {collaboration} {LIGO Scientific
  Collaboration and Virgo Collaboration}),\ }\href {\doibase
  http://arxiv.org/abs/1304.0670} {\bibfield  {journal} {\bibinfo  {journal}
  {arXiv}\ }\textbf {\bibinfo {volume} {1304}},\ \bibinfo {pages} {0670}
  (\bibinfo {year} {2013})}\BibitemShut {NoStop}%
\bibitem [{\citenamefont {{Aso}}\ \emph {et~al.}(2013)\citenamefont {{Aso}},
  \citenamefont {{Michimura}}, \citenamefont {{Somiya}}, \citenamefont
  {{Ando}}, \citenamefont {{Miyakawa}}, \citenamefont {{Sekiguchi}},
  \citenamefont {{Tatsumi}},\ and\ \citenamefont
  {{Yamamoto}}}]{2013PhRvD..88d3007A}%
  \BibitemOpen
  \bibfield  {author} {\bibinfo {author} {\bibfnamefont {Y.}~\bibnamefont
  {{Aso}}}, \bibinfo {author} {\bibfnamefont {Y.}~\bibnamefont {{Michimura}}},
  \bibinfo {author} {\bibfnamefont {K.}~\bibnamefont {{Somiya}}}, \bibinfo
  {author} {\bibfnamefont {M.}~\bibnamefont {{Ando}}}, \bibinfo {author}
  {\bibfnamefont {O.}~\bibnamefont {{Miyakawa}}}, \bibinfo {author}
  {\bibfnamefont {T.}~\bibnamefont {{Sekiguchi}}}, \bibinfo {author}
  {\bibfnamefont {D.}~\bibnamefont {{Tatsumi}}}, \ and\ \bibinfo {author}
  {\bibfnamefont {H.}~\bibnamefont {{Yamamoto}}},\ }\href {\doibase
  10.1103/PhysRevD.88.043007} {\bibfield  {journal} {\bibinfo  {journal}
  {\prd}\ }\textbf {\bibinfo {volume} {88}},\ \bibinfo {eid} {043007} (\bibinfo
  {year} {2013})},\ \Eprint {http://arxiv.org/abs/1306.6747} {arXiv:1306.6747
  [gr-qc]} \BibitemShut {NoStop}%
\bibitem [{\citenamefont {et~al.}(2011)}]{Indigo}%
  \BibitemOpen
  \bibfield  {author} {\bibinfo {author} {\bibfnamefont {I.~B.}\ \bibnamefont
  {et~al.}},\ }\href {\doibase https://dcc.ligo.org/LIGO-M1100296/public} {\
  (\bibinfo {year} {2011}),\
  https://dcc.ligo.org/LIGO-M1100296/public}\BibitemShut {NoStop}%
\bibitem [{\citenamefont {{Allen}}\ and\ \citenamefont
  {{Romano}}(1999)}]{1999PhRvD..59j2001A}%
  \BibitemOpen
  \bibfield  {author} {\bibinfo {author} {\bibfnamefont {B.}~\bibnamefont
  {{Allen}}}\ and\ \bibinfo {author} {\bibfnamefont {J.~D.}\ \bibnamefont
  {{Romano}}},\ }\href {\doibase 10.1103/PhysRevD.59.102001} {\bibfield
  {journal} {\bibinfo  {journal} {\prd}\ }\textbf {\bibinfo {volume} {59}},\
  \bibinfo {eid} {102001} (\bibinfo {year} {1999})}\BibitemShut {NoStop}%
\bibitem [{\citenamefont {Ade}\ \emph {et~al.}(2015)\citenamefont {Ade} \emph
  {et~al.}}]{Ade:2015xua}%
  \BibitemOpen
  \bibfield  {author} {\bibinfo {author} {\bibfnamefont {P.~A.~R.}\
  \bibnamefont {Ade}} \emph {et~al.} (\bibinfo {collaboration} {Planck}),\
  }\href@noop {} {\  (\bibinfo {year} {2015})},\ \Eprint
  {http://arxiv.org/abs/1502.01589} {arXiv:1502.01589 [astro-ph.CO]}
  \BibitemShut {NoStop}%
\bibitem [{\citenamefont {{Dominik}}\ \emph {et~al.}(2015)\citenamefont
  {{Dominik}}, \citenamefont {{Berti}}, \citenamefont {{O'Shaughnessy}},
  \citenamefont {{Mandel}}, \citenamefont {{Belczynski}}, \citenamefont
  {{Fryer}}, \citenamefont {{Holz}}, \citenamefont {{Bulik}},\ and\
  \citenamefont {{Pannarale}}}]{2015ApJ...806..263D}%
  \BibitemOpen
  \bibfield  {author} {\bibinfo {author} {\bibfnamefont {M.}~\bibnamefont
  {{Dominik}}}, \bibinfo {author} {\bibfnamefont {E.}~\bibnamefont {{Berti}}},
  \bibinfo {author} {\bibfnamefont {R.}~\bibnamefont {{O'Shaughnessy}}},
  \bibinfo {author} {\bibfnamefont {I.}~\bibnamefont {{Mandel}}}, \bibinfo
  {author} {\bibfnamefont {K.}~\bibnamefont {{Belczynski}}}, \bibinfo {author}
  {\bibfnamefont {C.}~\bibnamefont {{Fryer}}}, \bibinfo {author} {\bibfnamefont
  {D.~E.}\ \bibnamefont {{Holz}}}, \bibinfo {author} {\bibfnamefont
  {T.}~\bibnamefont {{Bulik}}}, \ and\ \bibinfo {author} {\bibfnamefont
  {F.}~\bibnamefont {{Pannarale}}},\ }\href {\doibase
  10.1088/0004-637X/806/2/263} {\bibfield  {journal} {\bibinfo  {journal}
  {\apj}\ }\textbf {\bibinfo {volume} {806}},\ \bibinfo {eid} {263} (\bibinfo
  {year} {2015})},\ \Eprint {http://arxiv.org/abs/1405.7016} {arXiv:1405.7016
  [astro-ph.HE]} \BibitemShut {NoStop}%
\bibitem [{\citenamefont {{Vitale}}\ and\ \citenamefont
  {{Evans}}(2016)}]{2016arXiv161006917V}%
  \BibitemOpen
  \bibfield  {author} {\bibinfo {author} {\bibfnamefont {S.}~\bibnamefont
  {{Vitale}}}\ and\ \bibinfo {author} {\bibfnamefont {M.}~\bibnamefont
  {{Evans}}},\ }\href@noop {} {\bibfield  {journal} {\bibinfo  {journal} {ArXiv
  e-prints}\ } (\bibinfo {year} {2016})},\ \Eprint
  {http://arxiv.org/abs/1610.06917} {arXiv:1610.06917 [gr-qc]} \BibitemShut
  {NoStop}%
\bibitem [{\citenamefont {Umst\"atter}\ \emph {et~al.}(2005)\citenamefont
  {Umst\"atter}, \citenamefont {Christensen}, \citenamefont {Hendry},
  \citenamefont {Meyer}, \citenamefont {Simha}, \citenamefont {Veitch},
  \citenamefont {Vigeland},\ and\ \citenamefont {Woan}}]{PhysRevD.72.022001}%
  \BibitemOpen
  \bibfield  {author} {\bibinfo {author} {\bibfnamefont {R.}~\bibnamefont
  {Umst\"atter}}, \bibinfo {author} {\bibfnamefont {N.}~\bibnamefont
  {Christensen}}, \bibinfo {author} {\bibfnamefont {M.}~\bibnamefont {Hendry}},
  \bibinfo {author} {\bibfnamefont {R.}~\bibnamefont {Meyer}}, \bibinfo
  {author} {\bibfnamefont {V.}~\bibnamefont {Simha}}, \bibinfo {author}
  {\bibfnamefont {J.}~\bibnamefont {Veitch}}, \bibinfo {author} {\bibfnamefont
  {S.}~\bibnamefont {Vigeland}}, \ and\ \bibinfo {author} {\bibfnamefont
  {G.}~\bibnamefont {Woan}},\ }\href {\doibase 10.1103/PhysRevD.72.022001}
  {\bibfield  {journal} {\bibinfo  {journal} {Phys. Rev. D}\ }\textbf {\bibinfo
  {volume} {72}},\ \bibinfo {pages} {022001} (\bibinfo {year}
  {2005})}\BibitemShut {NoStop}%
\bibitem [{\citenamefont {{Bin{\'e}truy}}\ \emph {et~al.}(2012)\citenamefont
  {{Bin{\'e}truy}}, \citenamefont {{Boh{\'e}}}, \citenamefont {{Caprini}},\
  and\ \citenamefont {{Dufaux}}}]{2012JCAP...06..027B}%
  \BibitemOpen
  \bibfield  {author} {\bibinfo {author} {\bibfnamefont {P.}~\bibnamefont
  {{Bin{\'e}truy}}}, \bibinfo {author} {\bibfnamefont {A.}~\bibnamefont
  {{Boh{\'e}}}}, \bibinfo {author} {\bibfnamefont {C.}~\bibnamefont
  {{Caprini}}}, \ and\ \bibinfo {author} {\bibfnamefont {J.-F.}\ \bibnamefont
  {{Dufaux}}},\ }\href {\doibase 10.1088/1475-7516/2012/06/027} {\bibfield
  {journal} {\bibinfo  {journal} {Journal of Cosmology and Astroparticle
  Physics}\ }\textbf {\bibinfo {volume} {6}},\ \bibinfo {eid} {027} (\bibinfo
  {year} {2012})},\ \Eprint {http://arxiv.org/abs/1201.0983} {arXiv:1201.0983
  [gr-qc]} \BibitemShut {NoStop}%
\end{thebibliography}%

\end{document}